\documentclass[12pt]{article}
\usepackage[margin=0.95 in]{geometry} 
\usepackage{amssymb,amsfonts}
\usepackage[all]{xy}
\usepackage{graphicx}
\usepackage{makecell}
\usepackage[utf8x]{inputenc}
\usepackage{amsmath}
\usepackage{array}
\usepackage{tikz}
\usepackage{tikz-cd}
\usepackage{mathtools}
\usepackage{mathrsfs} 
\usepackage[hidelinks]{hyperref}
\usepackage{cite}
\usepackage{amsmath} 
\usepackage{floatrow}
\usepackage{slashed}
\usepackage{dsfont}
\usepackage[all]{xy}
\usepackage{graphicx}
\usepackage{amsmath}
\usepackage{amssymb}
\usepackage{float}
\usepackage{array}
\usepackage{tikz}
\usetikzlibrary{calc}
\usepackage{mathtools}
\usepackage{mathrsfs} 
\usepackage[hidelinks]{hyperref}
\usepackage{cite}
\usepackage{tikz}
\usepackage{array, multirow}  
\numberwithin{equation}{section}
\newcommand{\be}{\begin{equation}}
\newcommand{\ee}{\end{equation}}

\def\bea{\begin{eqnarray}}
\def\eea{\end{eqnarray}}
\numberwithin{equation}{section}
\numberwithin{table}{section}

\begin{document}

\hypersetup{pageanchor=false}
\begin{titlepage}
\vbox{\halign{#\hfil    \cr}}  
\vspace*{15mm}
\begin{center}
{\Large \bf Equivariant Interpolations in Topological Holography
}

\vspace*{10mm} 

{\large
Jan Troost}
\vspace*{8mm}

Laboratoire de Physique de l'\'Ecole Normale Sup\'erieure \\ 
 \hskip -.05cm
 CNRS, ENS, Universit\'e PSL,  Sorbonne Universit\'e, \\
 Universit\'e  Paris Cit\'e 
 \hskip -.05cm F-75005 Paris, France	 

\vspace*{0.8cm}
\end{center}

\begin{abstract} {
We revisit equivariant Gromov-Witten theories on  $\mathbb{P}^1$ and on $\mathbb{P}^1 \times \mathbb{C}^2$. One can introduce three equivariant parameters associated to  rotations of the sphere as well as the two complex planes. A number of points in the parameter space  have known  holographic duals. These include the symmetric orbifold point dual to the  $AdS_3 \times S^3 \times \mathbb{C}^2$ string theory at string scale radius of curvature, the grand canonical Hurwitz theory and the product of two Kontsevich models. Within this framework, we discuss interpolations in the equivariant parameters. 
Firstly, we move between the small and large equivariant parameter regimes in Gromov-Witten theory on $\mathbb{P}^1$. At large equivariant parameter, the model is dominated by the pure topological gravity theories at the two fixed points while at small equivariant parameter the theory is equivalent to the grand canonical Hurwitz theory. 
The deformation is a solvable analogue for the interpolation in the transposition coupling in the moduli space of  the $AdS_3/CFT_2$ duality.   Moreover, we propose that the full equivariant correspondence between the Gromov-Witten theory on $\mathbb{P}^1 \times \mathbb{C}^2$ and the symmetric orbifold of the equivariant plane can be embedded in string theory. On the boundary side of that correspondence, we analyze the scaling limit from the equivariant to the ordinary cohomology ring for the Hilbert scheme of points on the plane in terms of Jack symmetric polynomials.  
We explicitly compute several structure constants of the equivariant cohomology ring and point out their  intriguing positivity and integrality properties.   
  }  
\end{abstract}

\end{titlepage}

\hypersetup{pageanchor=true}

\setcounter{tocdepth}{2}
\tableofcontents

\section{Introduction}
The quest to understand the holographic nature of quantum gravity \cite{tHooft:1993dmi,Susskind:1994vu} remains  open. Important clues are provided by the holographic nature of the Bekenstein-Hawking black hole entropy as well as by pairs of theories that are in holographic correspondence. The most convincing examples of these are either low dimensional theories of gravity or pairs that arise from the $AdS_{d+1}/CFT_d$ correspondence \cite{Maldacena:1997re}. 

The holographic pairs provided by string theory include bulk backgrounds where the non-zero fields live purely in the Neveu-Schwarz Neveu-Schwarz sector, which makes these theories amenable to study with standard world sheet techniques. A set of such backgrounds correspond to $AdS_3/CFT_2$ dualities. We can furthermore concentrate on an $AdS_3$ background with a radius of curvature equal to the string length $\sqrt{\alpha'}$ where $2 \pi \alpha'$  is the inverse fundamental string tension. In this case, many elements of the proof of the equivalence between the string theory in the bulk and the symmetric orbifold theory on the boundary have been put in place \cite{Eberhardt:2019ywk} -- the finalization and exploitation of the equivalence remain an active field of research. 

In this paper, we include the $AdS_3 \times S^3 \times \mathbb{C}^2$ model at string radius as one point of a proposed bulk landscape of  holographic duals. We review that the topological symmetric orbifold of $\mathbb{C}^2$ is dual to a Gromov-Witten theory on $\mathbb{P}^1 \times \mathbb{C}^2$. We interpret the duality as a topological twist of an $AdS_3/CFT_2$ correspondence \cite{Li:2020zwo}. We then place this topological dual pair into a landscape of equivariantly deformed Gromov-Witten theories on $\mathbb{P}^1_t \times \mathbb{C}^2_{t_1,t_2}$ that have been carefully studied in  mathematics
\cite{BryanGraber,BryanPandharipande,OPQuantumCohomology,OP1,OP2,OP3}. 

Within this global framework, our plan unfolds as follows.
In section \ref{EquivariantLandscape}, we paint the landscape and point to corners of the landscape in which we understand the holographic dual gauge theories. We will then be inspired by the expanse to study various interpolations in the equivariant couplings in order to probe the physics of the deformed theories. The rest of the paper splits into two parts in which we study two distinct interpolations. Firstly, in sections \ref{EquivariantGWP}, \ref{LargeEquivariance} and \ref{Correlators} we study the equivariant deformation of the sphere and compute observables that give insight into the physics at large equivariance. We demonstrate that the model interpolates between, on the one hand, two decoupled models of pure topological gravity and, on the other hand, the grand canonical Hurwitz theory. In section \ref{NearNonEquivariant} we analyze the theory at small equivariance. Secondly, in  section \ref{EquivariantHilbertScheme}, we study the boundary topological symmetric orbifold model with equivariant parameters $(t_1,t_2)$ in its Hilbert scheme incarnation. We provide insight into the necessary mathematical background and  analyze the limit in which we recuperate the standard cohomology of the symmetric orbifold. Moreover, we compute various structure constants of the operator algebra explicitly and recall their intriguing properties, which have only partially been proven.  The detailed treatments in sections \ref{EquivariantGWP} to \ref{EquivariantHilbertScheme} are but two case studies in the large landscape sketched in section \ref{EquivariantLandscape}. We conclude in section \ref{Conclusions} with a discussion of further questions that our detailed studies, as well as the broader framework, raise.

\section{The Equivariant Landscape}
\label{EquivariantLandscape}
In this section, we briefly review the landscape of mathematical equivalences that form the backdrop to our exploration. We recall  dualities between gravitational and gauge theories that have been lifted to  theorems \cite{BryanGraber,BryanPandharipande,OPQuantumCohomology,OP1,OP2,OP3}, suggest how to embed them in string theory and point out special cases that have been discussed in some detail in the literature. Thus, we set the stage for the detailed interpolations in the rest of the paper. 
\subsection{The First Equivalence Theorem}
Consider the manifold $\mathbb{P}^1 \times \mathbb{C}^2$.
We can define three torus actions on the various factors. Firstly, a torus action on the $\mathbb{P}^1$ acts by multiplication near zero and the inverse multiplication near infinity. We associate the equivariant parameter $t$ to this torus action. Secondly, we have two torus actions on the two $\mathbb{C}$ factors with equivariant parameters $(t_1,t_2)$. 

In a first instance, we consider the theory on $\mathbb{P}^1 \times \mathbb{C}^2$ with equivariant parameters along the complex plane only. It has been proven that the equivariant Gromov-Witten theory on this bulk theory is equivalent to the equivariant  quantum cohomology  of the symmetric orbifold of the space $\mathbb{C}^2$, which in turn is equivalent to the quantum cohomology of the Hilbert scheme of points \cite{BryanGraber,BryanPandharipande,OPQuantumCohomology}. The correspondence identifies the string coupling $u$ of the Gromov-Witten theory with the quantum cohomology parameter $q$ in the boundary with a relation $q=-e^{iu}$.  
The proof of the equivalence proceeds in two steps. Firstly, the relative three-point functions on both sides of the correspondence for the multiplication by the transposition are proven to correspond to multiplication by the operator $M_2$ \cite{BryanPandharipande}:
\begin{align}
-M_2(q,t_1,t_2) & = \frac{t_1+t_2}{2} \sum_{k>0} \left( k \frac{(-q)^k+1}{(-q)^k-1}-\frac{(-q)+1}{(-q)-1} \right) \alpha_{-k} \alpha_k \\
&+ \frac{1}{2} \sum_{k,l>0} (t_1 t_2 \alpha_{k+l} \alpha_{-l} \alpha_{-k} - \alpha_{-k-l} \alpha_k \alpha_l) \, ,
\label{M2Quantum}
\end{align}
acting in a chiral bosonic Fock space (with operators $\alpha_n$) corresponding to the space of operators in the topological theory. 
Secondly, it is proven that the full theory is determined by this operator. This is because, on the one hand, the transposition (or its dual) generates the full set of operators (just as it generates all elements in the symmetric group), and on the other, by an argument of semi-simplicity (based on the fact that for generic parameters, the eigenvalues of $M_2$ are distinct while no singularities arise for the limiting values). The equivalence theorem follows from these observations \cite{BryanGraber,BryanPandharipande,OPQuantumCohomology}. It states that:
\begin{equation}
GW(\mu,\nu,\rho; q,t_1,t_2) = \langle \mu,\nu,\rho \rangle_{q,t_1,t_2} \, , \label{Theorem1}
\end{equation}
namely that the string three-point functions $GW$ in the Gromov-Witten theory for relative insertions determined by the partitions $(\mu,\nu,\rho)$ agree with the quantum cohomology structure constants $\langle \mu,\nu,\rho \rangle$ in the symmetric orbifold on the equivariantly deformed complex two-plane.

It will be interesting for us to consider a rescaling of all annihilation operators $\alpha_{k>0}$ by a factor of $t_1$ and  creation operators $\alpha_{-k}$ by the inverse factor, which gives rise to the alternative expression for the operator $\tilde{M}_2$:
\begin{align}
-\tilde{M}_2(q,t_1,t_2) &= \frac{t_1}{2} \Big((1+\frac{t_2}{t_1}) \sum_{k>0} \left( k \frac{(-q)^k+1}{(-q)^k-1}-\frac{(-q)+1}{(-q)-1} \right)   \alpha_{-k} \alpha_k 
\nonumber \\
& + \sum_{k,l>0} (\frac{t_2}{t_1} \alpha_{k+l} \alpha_{-l} \alpha_{-k} -  \alpha_{-k-l} \alpha_k \alpha_l) \Big)\, .  \label{M2TildeQuantum}
\end{align}
In this paper, we will be interested in the classical cohomology  and we will set $q=-1$. In this manner, we reduce the boundary theory to a standard topological quantum field theory entirely determined by a commutative Frobenius algebra and its structure constants.  The operators simplify to the cubic expressions:
\begin{align}
-M_2(t_1,t_2) & = \frac{1}{2} \sum_{k,l>0} (t_1 t_2 \alpha_{k+l} \alpha_{-l} \alpha_{-k} - \alpha_{-k-l} \alpha_k \alpha_l) \, 
\label{M2} 
\end{align}
and
\begin{align}
-\tilde{M}_2(t_1,t_2) &= \frac{t_1}{2} \sum_{k,l>0} (\frac{t_2}{t_1} \alpha_{k+l} \alpha_{-l} \alpha_{-k} -  \alpha_{-k-l} \alpha_k \alpha_l) \Big)\, . \label{M2Tilde}
\end{align}
Note that for the anti-diagonal parameter slice $t_1=-t_2$, the quantum cohomology corrections in the rescaled operator 
$\tilde{M}_2$ in equation (\ref{M2TildeQuantum}) are  absent.
\begin{figure}[ht]
\includegraphics[width=8cm]{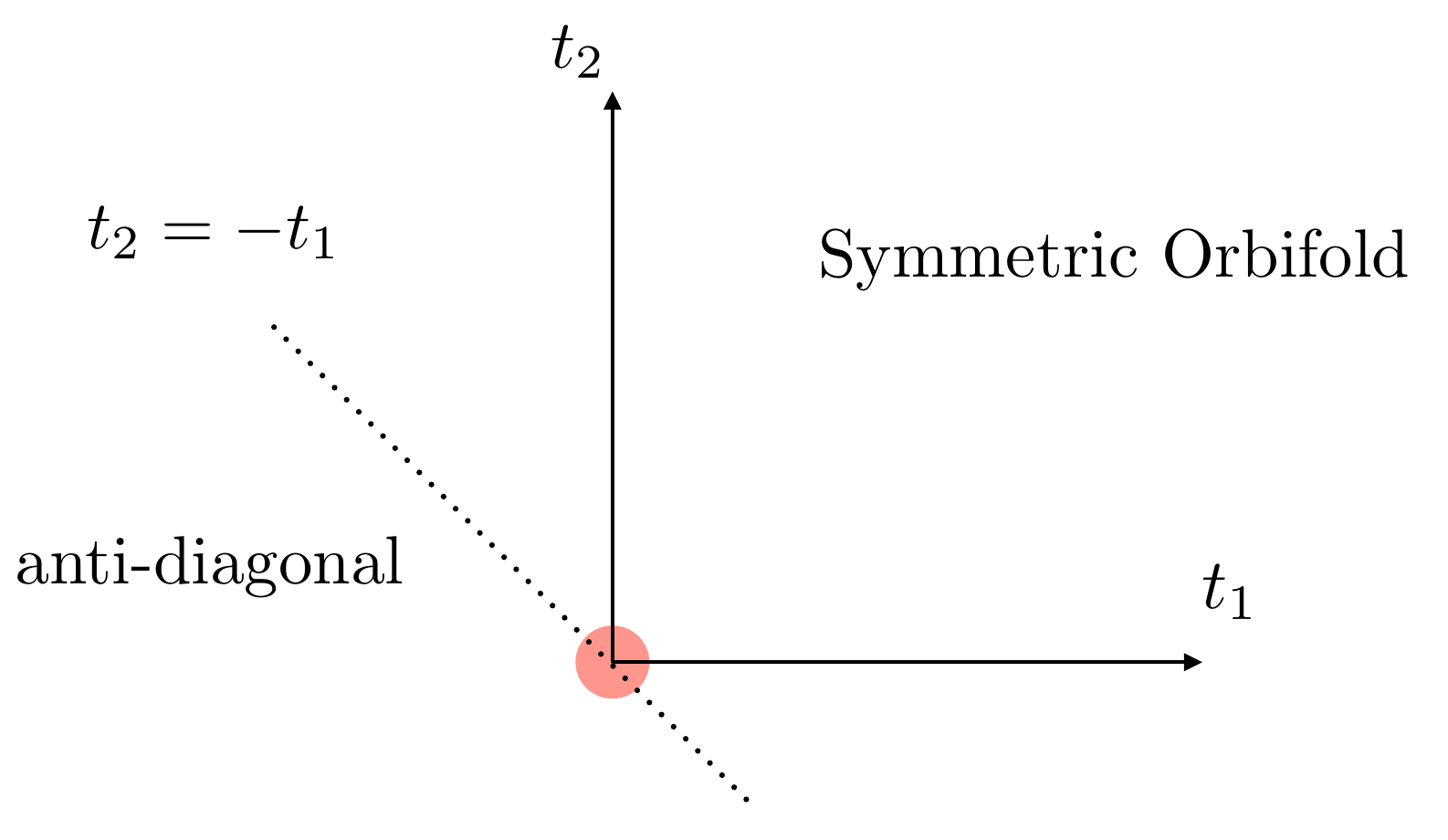}
\caption{There is a two-parameter space of equivariant symmetric orbifolds. The ordinary symmetric orbifold lies at the origin. The anti-diagonal models are classical.}
\label{EquivariantOrbifoldParameterSpace}
\end{figure}

\subsubsection{The AdS/CFT Correspondence}
Let us link these models to more familiar theories. 
Firstly, we note that when the equivariant parameters are zero, the operator $M_2$ reduces to the pure join operator:
\begin{equation}
M_2 = \frac{1}{2} \sum_{k,l>0} \alpha_{-k-l} \alpha_k \alpha_l \, .
\end{equation}
This joining interaction corresponds to (the action of the transposition in) the  topological symmetric orbifold  on the plane \cite{Li:2020zwo} or the cohomology of the Hilbert scheme of points on the plane \cite{LS}. In \cite{Li:2020zwo}, this model was pointed out to be a closed topological subsector of the $AdS_3/CFT_2$ correspondence, and a treasure trove of correlation functions were computed \cite{Ashok:2023mow}, which among other results, proved a conjecture in \cite{Pakman:2009ab}. Moreover, we have strong evidence for the duality between the $AdS_3 \times S^3 \times M_4$ string theory at string radius and the  symmetric orbifold of the  manifold  $M_4$ \cite{Eberhardt:2019ywk}. A topological subsector of these correspondences therefore reads:
\begin{equation}
(AdS_3 \times S^3)_{R=\sqrt{\alpha'}} \times \mathbb{C}^2 \leftrightarrow Sym_N(\mathbb{C}^2) \leftrightarrow GW(\mathbb{P}^1 \times \mathbb{C}^2) \, .
\end{equation}
The right hand side string theory must therefore be a topologically twisted version of the left hand string theory  \cite{Li:2020zwo}. It is natural to interpret the space $\mathbb{P}^1$ as the boundary of the euclidean $AdS_3$ space. 
The matching of the three-point functions in $AdS_3/CFT_2$
\cite{Gaberdiel:2007vu,Dabholkar:2007ey} has a topological counterpart in the theorem (\ref{Theorem1}).

\subsubsection{Towards a String Theory Embedding}
The theorem (\ref{Theorem1}) is valid for a two-dimensional space of equivariant parameters. It would be interesting to recuperate the full space of parameters in string theory. One way to embed the duality in string theory would be to start from a D1-D5 system in flat space along the directions $0,1$ and $6,7,8,9$.  The latter represent the complex plane $\mathbb{C}^2$. Crucially, one must render the theory on the D5-brane equivariant by introducing an $\Omega$ background \cite{Nekrasov:2002qd,Nekrasov:2010ka} for the D5-brane and the gauge theory living on it. Then, if we have $N$ D1-branes inside one D5-brane and we are on the Higgs branch of the (non-commutative) gauge theory on the D5-brane,  the D1-branes represent $N$ interchangeable instantons whose geometry is captured by the equivariant symmetric orbifold or its Hilbert scheme resolution. After S-duality, in the presence of a single dual NS5-brane, one may hope to recuperate the desired Gromov-Witten theory after a topological twist.  It would be very interesting to prove the viability of the string theory embedding -- the theorem (\ref{Theorem1}) about the topological holographic duality is valid independently.

\subsubsection{
The Anti-diagonal Model}
Note that after the rescaling, the anti-diagonal model (with $t_2=-t_1$) has an $\tilde{M}_2$ operator (\ref{M2TildeQuantum}) which coincides with the standard action of the transposition in Hurwitz theory \cite{OP1,OP2,OP3}  (with coupling $t_1$). The quantum anti-diagonal model was related to $q$-deformed Yang-Mills theory in \cite{Aganagic:2004js}.

\subsection{A Second Duality}
A second duality that we will explore is the equivariant Gromov-Witten theory on $\mathbb{P}^1$. A  theorem states that the non-equivariant Gromov-Witten theory on $\mathbb{P}^1$ is equivalent to a topological quantum field theory that we refer to as the (grand canonical) Hurwitz theory. It is a direct sum of $S_N$ Dijkgraaf-Witten theories in two dimensions. We have the equivalence \cite{OP1}:
\begin{equation}
GW (\mathbb{P}^1) \leftrightarrow \text{Hurwitz Theory}
\, \,  \label{Theorem2}
\end{equation}
Again, this can be viewed as a proven string theory/gauge theory duality \cite{Li:2020zwo,Benizri:2024mpx}. The Gromov-Witten theory can be equivariantly deformed, and we can ask whether the holographic duality can also be reformulated along the whole line. In the following sections, we explore the theory both at large and small equivariance.\footnote{It would be good to connect this model precisely to the equivariant theory on $\mathbb{P}^1_t \times \mathbb{C}^2_{t_1,t_2}$. We offer  a few comments. 
We note that in the equivariant theory on $\mathbb{P}^1 \times \mathbb{C}^2$, at $t_2=-t_1$, the operator $\tilde{M}_2$ (\ref{M2TildeQuantum}) matches the operator ${\cal F}_2$ in Hurwitz theory \cite{OP1} with a map of couplings $t_1 \leftrightarrow g_s/t$ where $g_s$ is the string coupling in the Gromov-Witten theory and $t$ is the equivariant parameter on the $\mathbb{P}^1$. At small equivariance $t$ and fixed string coupling, the $t_1$ parameter is large and the equivariant theory on  the $\mathbb{C}^2$ plane  localizes to a point - the parameter $g_s$ remains an independent parameter.}
\begin{figure}[ht]
\includegraphics[width=8cm]{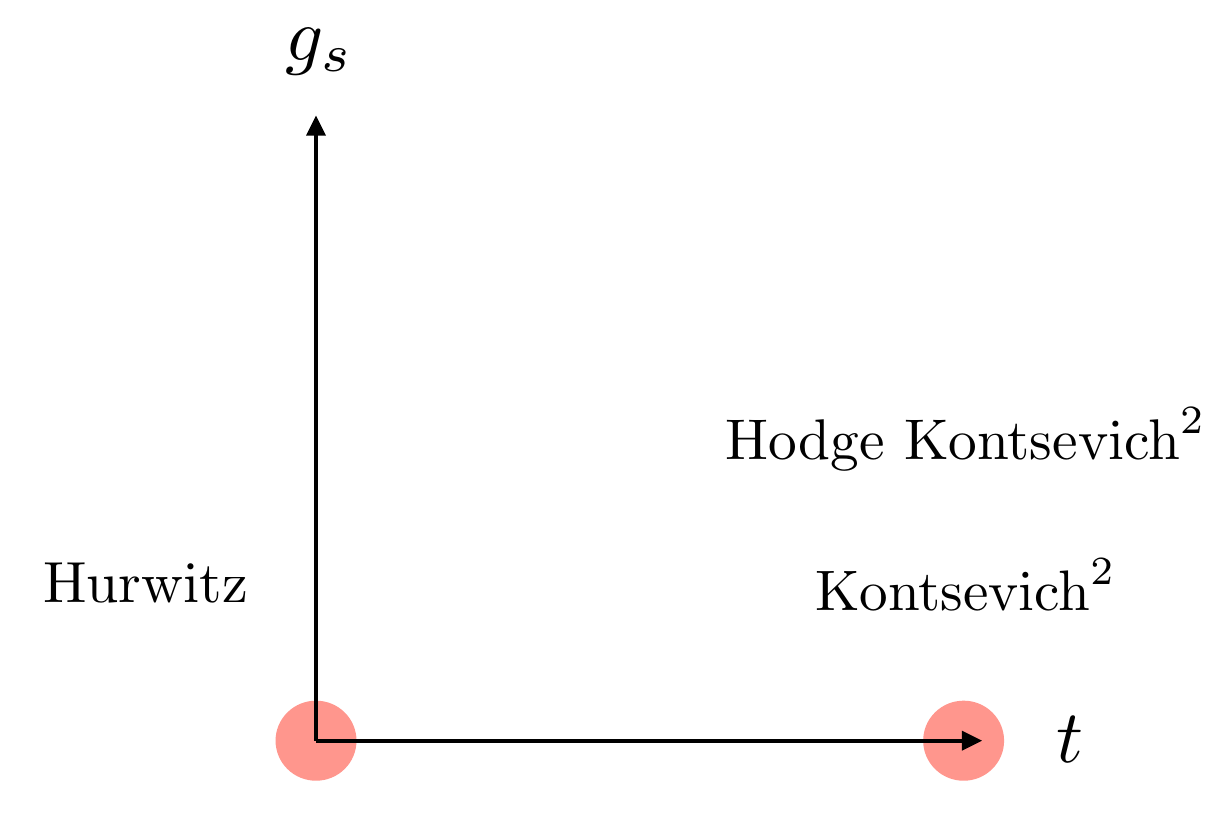}
\caption{The equivariant Gromov-Witten model on the sphere has two parameters. At zero equivariance, we have a Hurwitz dual while at infinite equivariance one finds the product of two Kontsevich models. Turning on the string coupling in the latter introduces Hodge class corrections to the topological gravity models.}
\label{EquivariantP1ParameterSpace}
\end{figure}
Note that while the duality (\ref{Theorem1}) is based on a match of relative invariants to three-point functions, the duality (\ref{Theorem2}) is formulated as an exact map between the algebras and the expectation values of operators \cite{OP1}.

\subsection{A Remark on  Charges}
It may be informative to pause and identify a few charges in the various models. 
\subsubsection{The Virasoro Degree}
Let us assign various degrees. Firstly, we introduce the Virasoro $L_0$ degree which equals $n$ for the operator $\alpha_{-n}$.
The operator $M_2(q,t_1,t_2)$ then commutes with the Virasoro degree, which is therefore preserved by the transposition interaction that has  Virasoro degree  zero: 
\begin{align}
[L_0, \alpha_{-n}] &= n \, \alpha_{-n}
\nonumber \\
[L_0,M_2] &= 0 
\, .
\end{align}
The equivariant parameters also have Virasoro degree zero. 
\subsubsection{The R-Charge}
Secondly, at the symmetric orbifold point, or in the Hilbert scheme of points on $\mathbb{C}^2$, we have zero equivariant parameters $t_i$ in the classical operator $M_2(t_1,t_2)$. In this theory, we introduce a charge $R$ such that the R-charge of $\alpha_{-n}$ is $n-1$ \cite{Li:2020zwo} and we therefore have that the transposition operator, which is a sum of two-cycles, has charge one:
\begin{align}
[R,\alpha_{-n}] &= (n-1) \, \alpha_{-n}
\nonumber \\
[R,M_2(0,0)] &=  M_2 (0,0) \, .
\end{align}
%
Next, we turn on the parameters $t_i$. 
Suppose we  want the full $M_2(q,t_1,t_2)$ operator to continue to have charge $+1$. We must then assign R-charge $1$ to each equivariant parameter $t_i$. The first and second term in (\ref{M2Quantum}) then also have R-charge one since the oscillator charges cancel in the first term and add to minus one in the second:
\begin{align}
[R,t_i] &= t_i
\nonumber \\
[R,M_2(q,t_1,t_2)] &= M_2(q,t_1,t_2) \, .
\end{align}
Note that the rescaling operation to go from the operator $M_2$ to the operator $\tilde{M}_2$ is such that the R-charges of the new oscillators become equal to their Virasoro degree. The only difference between the grading of the operator $\tilde{M}_2$ then lies in assigning Virasoro degree $0$ to the coupling $t_1$ up front or, alternatively, assigning it R-charge  $1$.

\subsection{Exploring Corners of the Parameter Space}
The rest of the paper consists of explorations of the corners of the parameter space that we presented -- see Figures \ref{EquivariantOrbifoldParameterSpace} and \ref{EquivariantP1ParameterSpace}. In the following sections, we study the equivariant Gromov-Witten theory on $\mathbb{P}^1$. We identify the holographic dual theory at infinite equivariance parameter and study some of the features of perturbation theory around the points at infinite coupling and at zero coupling, both of which have identifiable field theory duals. In section \ref{EquivariantHilbertScheme} on the other hand, we explore the topological theory on the equivariant Hilbert scheme. In particular, we unpack the mathematical description of  the equivariant cohomology, obtain a hands-on understanding of the structure constants of the associated ring, and identify the scaling limit that allows us to make contact with the classical cohomology of the Hilbert scheme of points on the plane (which is equivalent to the topological symmetric orbifold). A common  theme is  the equivariant interpolation around and between identifiable holographic pairs. 

\section{The Equivariant Gromov-Witten Theory}
\label{EquivariantGWP}
The Gromov-Witten theory of curves was solved in the triptych \cite{OP1,OP2,OP3} through  equivariant localization and degeneration. One outcome was the oscillator expression for the equivariant Gromov-Witten invariants on the sphere.
The equivariant parameter for Gromov-Witten theory on the sphere controls a perturbation theory in the transposition operator. Here, we explicitly compute the equivariant correlator expressions \cite{OP1,OP2,OP3}  in order to analyze how they depend on the parameter that interpolates between weak and strong coupling in the transposition operator. Through the calculation of observables, we confirm the intuitive picture in which we expect disconnected covers to dominate the observables at weak coupling while connected covers determine the expectation values at a point dual to the Hurwitz theory. We note that in $AdS_3/CFT_2$, the transposition interaction has been argued to link the bulk point in the moduli space with the boundary symmetric orbifold point -- see \cite{Eberhardt:2021vsx} and references therein. The model we have at hand here is a theory in which one can exactly solve the interpolation in the transposition operator between the weak and strong coupling regime. 

Since our analysis is lengthy, we divide it up as follows. 
In this section, we briefly review results on the equivariant Gromov-Witten correlators on $\mathbb{P}^1$ \cite{OP2}. We discuss the degeneration formula as well as the oscillator expression for the correlators. For more background, we refer to the original papers \cite{OP1,OP2,OP3} as well as \cite{Troost:2026pmx}. 
We analyze the theory at large equivariant parameter in section \ref{LargeEquivariance} and compute explicit observables in the large equivariance expansion in section \ref{Correlators}.  We find  characterizations of the  infinite equivariance theory and the first orders in perturbation theory.  Section \ref{NearNonEquivariant} sets up a strong coupling perturbation theory near the non-equivariant theory. The latter is known to correspond to a second quantized Hurwitz topological quantum field theory
with $S_N$ gauge symmetry \cite{OP1}. %

\subsection{The Localization and Degeneration}
The correlators of the equivariant Gromov-Witten theory on the sphere were calculated in \cite{OP2,OP3} in terms of handy oscillator expressions. They were determined by localization with respect to a rotation with two fixed points labeled $0$ and $\infty$. Localization is combined with degeneration in order to compute the contributions at the fixed points and glue them back together to form the full amplitude. The factors allow for oscillator expressions that can be simplified after gluing \cite{OP2,OP3}. 

The standard cohomology of $\mathbb{P}^1$ is generated by the identity and the volume class $\omega$. After introducing an equivariant parameter $t$ that weighs rotations near $0$ and $\infty$ oppositely, the equivariant cohomology can be written as linear combinations 
of 
 the Poincar\'e dual classes $[0]$ and $[\infty]$ to the fixed points.
For the equivariant descendant operators $\tau_k[1]$ and $\tau_l[\infty]$, the disconnected correlators  at  degree $d$ are captured by a generating function $G_d$:
\begin{align}
G_d(z_i,w_j,g_s,t) &= \sum_{g,k_i,l_j} g_s^{2-2g}  z_i^{k_i+1} w_j^{l_j+1} \langle \prod_{i=1}^n \tau_{k_i}[1]  \prod_{j=1}^m \tau_{l_j}[\infty] \rangle_{g,d} \, .
\label{G}
\end{align}
We denoted the genus counting parameter by the string coupling $g_s$ and introduced the variables $(z_i,w_j)$ that keep track of the descendant order of the insertions. The parameter $d$ is the degree of the cover of the sphere. 
Combining localization with degeneration gives rise to factorized moduli space integral contributions at the fixed points, which combine into the correlator expression:
\begin{align}
G_d(z_i,w_j,g_s,t) &=\frac{1}{z(\mu)} \sum_{|\mu|=d}
(\frac{g_s}{t})^{l(\mu)} (-\frac{g_s}{t})^{l(\mu)} 
t^{-d-n} (-t)^{-d-m} (\prod_i \frac{\mu_i^{\mu_i}}{\mu_i!})^2 
\nonumber \\
 & \qquad \qquad \qquad
H(\mu,tz,\frac{g_s}{t}) \times  H(\mu,-tw,- \frac{g_s}{t}) \label{Correlator}
\, .
\end{align}
 The normalization $z(\mu)= |\text{Aut} \, \mu |  \prod_{i=1}^{l(\mu)} \mu_i$ equals the number of automorphisms of the partition $\mu$ of the degree $d$ of the correlator times a factor associated to cyclic symmetry. The sum over partitions is  a sum over intermediate states. The central dynamical ingredients in the formula are the connected linear Hodge integrals $H^\circ$ which are integrals over compactified moduli spaces of Riemann surfaces $\overline{M}_{g,n}$:
\begin{equation}
H^{\circ}_g(z_i) = \prod z_i
\int_{\overline{M}_{g,n}} \frac{\Lambda_g}{\prod (1-z_i \psi_i)} \,  \label{LinearHodgeIntegrals}
\end{equation}
of the Hodge class $\Lambda_g=1-\lambda_1 +\lambda_2- \dots \pm \lambda_g$ and tangent space classes $\psi_i$. 
Their disconnected counterparts $H$ appearing in the correlator (\ref{Correlator})
 allow for an expression in terms of oscillators in a fermionic (or bosonic) Fock space.
 \subsection{The Correlators in Oscillator Form}
 The Hodge integrals $H$ are analogs of Hurwitz numbers and, as such, feature an exponential interaction $e^{\frac{g_s}{t} {\cal F}_2}$ \cite{OP2} of the transposition operator ${\cal F}_2$ whose role was reviewed  in \cite{Troost:2026pmx}. In the final oscillator expression for the correlators, these interaction terms are absorbed in the operator insertions \cite{OP2}
 \begin{equation}
A(tz,g_sz) = S(g_sz)^{tz} \sum_{k \in \mathbb{Z}} 
\frac{\zeta(g_sz)^k}{(tz+1)_k} {\cal E}_k(g_sz) \, ,
\end{equation}
which depend on functions
\begin{equation}
\zeta(z)= 2 \sinh \frac{z}{2} \, , \qquad S(z)=\frac{\zeta(z)}{z}
\, ,
\end{equation}
and operators ${\cal E}_k$ which are complicated fermion bilinear operators whose properties are reviewed in \cite{OP2,OP3,Troost:2026pmx}.
The final oscillator expression for the generating function $G(z,w)=\sum_d q^d G_d(z,w)$ becomes \cite{OP2}:
\begin{equation}
G(z,w) = \langle \prod_i \frac{A(tz_i,g_sz_i)}{g_s} e^{\alpha_1} (\frac{q}{g_s^2})^{\cal H} e^{\alpha_{-1}}  \prod_j \frac{A^\ast(-tw_j,g_sw_j)}{g_s} \rangle
\end{equation}
where $\alpha_1$ is the first oscillator of the bosonized fermion, $q$ is the parameter counting degree and the Hamiltonian ${\cal H}$ equals $d$ for each degree $d$ cover. 
%
%
Finally, there is a generalization of these results to relative Gromov-Witten theory \cite{OP3} of which we will only use  a simple case in the following. 

\subsection{The Global Properties}
All these results are parameterized by 
the string coupling $g_s$,  the degree counting parameter $q=e^{-a}$, where $a$ is a complexified area of the target sphere, and the equivariant parameter $t$.
A first crucial fact to keep in mind is that the operators are constructed by commuting the operator $e^{\frac{g_s}{t} {\cal F}_2}$ through oscillators, combined with analytic continuation \cite{OP2}. 
Secondly, at $t=0$, we have a gravity/gauge theory equivalence between (non-equivariant or ordinary) Gromov-Witten theory on the sphere  and the Hurwitz theory \cite{OP1}. In other words, at strong coupling $g_s/t$, we have a non-equivariant gravitational bulk theory dual to a grand canonical $S_N$ gauge theory \cite{OP1,Troost:2026pmx}.

\section{The Large Equivariance Theory}
\label{LargeEquivariance}

The appearance of the exponential interaction term in the transposition operator with coupling $g_s/t$ suggests that the model at large equivariance should drastically simplify and become free in an appropriate sense. Alternatively, a large equivariant parameter $t$ will project the theory onto a close neighborhood of the fixed points which likewise indicates a drastic  simplification. 
In this section,  we analyze the theory at large equivariant parameter $t$ and determine the scaling of the $(n,m)$-point functions and of the operators and we identify the dominant terms in amplitudes up to the second order in the $1/t$ expansion. 

\subsection{The Scaling of the Correlators}
To determine the scaling of the correlators in $(g_s,t)$, recall the degeneration formula:
\begin{align}
G_d(z_i,w_j,g_s) &=\frac{1}{z(\mu)} \sum_{|\mu|=d}
(\frac{g_s}{t})^{l(\mu)} (-\frac{g_s}{t})^{l(\mu)} 
t^{-d-n} (-t)^{-d-m} (\prod_i \frac{\mu_i^{\mu_i}}{\mu_i!})^2 
\nonumber \\
 & \qquad \qquad \qquad
H(\mu,tz,\frac{g_s}{t}) \times  H(\mu,-tw,- \frac{g_s}{t})
\, .
\label{Degeneration}
\end{align}
The total string coupling $g_s$ dependence is $g_s^{2g-2}$ where $g$ is the total genus of the covering surface.  We also know that the $z_i$ dependence of a $\tau_{k_i}$ descendant correlator is $z_i^{k_i+1}$. Using these two facts and the structure of the  formula (\ref{Degeneration}), we obtain that a $\prod_i \tau_{k_i}$ correlator scales as
$g_s^{2g-2} t^{-2d-2g+2+\sum_i k_i}$.
Since the non-equivariant limit $t \rightarrow 0$ is well-defined, the power of $t$ will always be non-negative. In the 
 $t \rightarrow 0$ limit, the power of $t$ is zero, which confirms the known dimension constraint \cite{OP1} 
\begin{equation}
2g-2+2d=\sum_i k_i \qquad \quad \text{at } \, t=0 \, ,
\label{NonEquivariantConstraint}
\end{equation}
on non-equivariant correlators.
In the general setting, we summarize the scaling behavior in terms of the power $n_t$ of the equivariant parameter $t$ which satisfies
\begin{equation}
2g-2+2d+ n_t = \sum_{i} k_i \, . \label{EquivariantConstraint}
\end{equation}
The differential degree of $t$ on the moduli space is effectively minus one. Because of the good behavior of the correlators in the non-equivariant limit, we have that $n_t \ge 0$. 

\subsection{The Rescaling of the Operators}
Let us assume we fix the external insertion descendant degrees $k_i$. Because of the scaling of the correlators with the  degrees $k_i$, we rescale the operators with a power of the equivariant parameter:
\begin{equation}
\tilde{\tau}_{k}(\omega) = t^{-k} \tau_k(\omega)
\, .
\end{equation}
As a consequence, the correlators have a behavior in $t$ that is dictated by the power:
\begin{equation}
\tilde{n}_t =2-2g-2d 
\, . 
\label{tScaling}
\end{equation}
We have the lower bound $\tilde{n}_t \ge - \sum k_i \, .$
There is also an upper bound related to the number of connected components of the covering surface. Indeed, the total genus $g$ of a disconnected contribution is:
\begin{equation}
g = \sum_i g_i - \# \text{components} +1
\, ,
\end{equation}
where $g_i$ are the genera of the connected components. 
The total genus $g$ has a (negative) lower bound in terms of the number of components of the surface.
\subsection{The Parameter Regime}
The equivariant Gromov-Witten invariants summed over the degree $d$ naively depend on three parameters, the instanton weight $q$, the string coupling $g_s$ and the equivariant parameter $t$. However, the constraint equation
(\ref{EquivariantConstraint})  implies that only two out of the three parameters are independent.\footnote{Recall that in the non-equivariant case, the constraint (\ref{NonEquivariantConstraint}) enforces that the correlators only depend on the ratio $q/g_s^2$. The equivariant parameter  decouples the world sheet instanton and the perturbative string expansion. 
For a given correlator, in the equivariant theory, at fixed $k_i$ and degree $d$, it is possible to obtain (a finite number of) string loop corrections. } One can identify these parameters as the ratio of the K\"ahler and equivariant parameters:
\begin{equation}
\tilde{q}= \frac{q}{t^2} \, ,
\end{equation}
as well as the redefined string coupling
\begin{equation}
\tilde{g_s} = \frac{g_s}{t} \, .
\end{equation}  
The large equivariant parameter limit we consider, is one in which we keep both the original instanton weight $q$ and the string coupling $g_s$ fixed. 

 \subsection{The Dominant Contributions}
We identify dominant connected contributions to the correlation functions. 
From the exponent (\ref{tScaling}) of the correlators in the parameter $t$ and the fact that for connected contributions $g \ge 0$, while all contributions have $d \ge 0$, it is manifest that $(g,d)=(0,0)$ is the dominant contribution.
It is moreover clear from the degeneration formula (\ref{Degeneration}) that at zero degree, all $(n,m)$ point functions factorize:
\begin{align}
G_0(z_i,w_j,g_s) &= 
t^{-n} (-t)^{-m} 
H(tz_i,\frac{g_s}{t}) \times  H(-tw_j,- \frac{g_s}{t})
\, .
\label{LeadingG}
\end{align} 
We therefore distinguish decoupled theories at the fixed points $0$ and $\infty$. At the fixed point $0$, we have the correlators:\footnote{The Gromov-Witten theory on the sphere has unstable correlators precisely when $d=0$. Their values are set by fiat for consistency with integrability \cite{OP2}. }
\begin{equation}
G_0 (z_i,g_s) = t^{-n} H(tz_i,\frac{g_s}{t}) \, .
\end{equation}
 All genus $0$ (connected) correlators correspond to genus $0$ ordinary (connected) intersection integrals on the moduli space of Riemann surfaces. These are the correlators of pure topological gravity or the Gromov-Witten theory of a point \cite{Witten:1989ig}. The total leading term (\ref{LeadingG}) is therefore the product of two point theories (at $0$ and $\infty$). This makes intuitive sense as the large equivariant parameter (ultra-)localizes the physics at the fixed points.  
 The connected genus zero contributions will therefore be genus zero contributions at one of the two fixed points, which reduce to integrals on the moduli space of punctured spheres, for which there is a closed formula. 
 The full leading result (\ref{LeadingG}) is given by the intersection theory of two points times an equivariant intersection product of classes responsible for the separate powers of the equivariant parameter $t$. 

Note that, in fact, the degree zero correlators factorize for {\em all} genera. We therefore identify the linear Hodge integrals (\ref{LinearHodgeIntegrals}) as the connected building block of all degree zero correlators. Starting at genus one, the Gromov-Witten theory of two points is augmented by non-trivial Hodge class contributions. 
For the Gromov-Witten theory at $\mathbb{P}^1$ at degree zero, the only connected contributions at the leading order arise when we take a one-sided correlator. Thus, we have a connected contribution which is the sum of two connected linear Hodge integral generators $H^\circ$. 

\subsection{The Twist and the Cover}
The subleading contributions correspond to the shifted power $\tilde{n}_t=0$ in the variable $t$ -- see equation (\ref{tScaling}) -- and arise from either $(g,d)=(1,0)$ or $(g,d)=(0,1)$. 
\subsubsection{Linear Hodge Integral Corrections at Higher Genus}
As we saw, the $(g,d)=(1,0)$ corrections are again factorized since they are of degree zero. From both fixed points, there are linear Hodge integral corrections with an integrand proportional to the Hodge class $\lambda_1$, as well as pure gravity amplitudes at genus one. It should be clear how this pattern continues with higher genus linear Hodge integrals as we study corrections of the type $(g,d)=(g,0)$. The corrections carry extra factors of $(g_s^2/t^2)^g$ compared to the genus zero amplitudes. They correspond to $g$ pairs of insertions of the interaction operator ${\cal F}_2$. All linear Hodge integrals are algorithmically computable \cite{Faber:1998gsw} and a number of closed formulas exist. 
In short, if  we set the parameter $q$ to zero, we obtain two point like theories dressed with higher genus linear Hodge integral corrections. 

\subsubsection{The Genus Zero Degree One Corrections}
The second type of sub-leading correction carries an extra  factor of $q/t^2$ and has genus zero and degree one. 
This is the leading contribution with a non-zero degree, and when exponentiated, it forcibly leads to the leading disconnected  contribution at any non-zero degree $d$.
The degree one  connected contributions introduce many types of extra non-zero amplitudes. Because we have an edge at a vertex -- see \cite{OP2,Troost:2026pmx} for an introduction to the Feynman diagrams — the associated $\psi$ tangent class insertion has an arbitrary power and can lead to a considerable number of extra non-vanishing amplitudes. 
We also keep in mind that the concept of a connected and stable contribution refers to the $\mathbb{P}^1$ theory.
Therefore, with an edge, we can  have stable connected contributions arising from unstable point vertices connected by an edge. Thus, we can have left and right low point functions that attach to an edge and become stable.  
Apart from the twist due to the Hodge class, at degree one, we have the first interactions between the two theories of a point.  We have connected contributions with both left and right insertions that extend the factorized point theories to a theory of topological gravity coupled to matter. 

\begin{figure}[H]

\begin{tikzpicture}[
  scale=1,
  vertex/.style={circle, fill=black, inner sep=1.5pt}
]


\node[vertex] (d1) at (0,0) {};
\node[vertex] (d1p) at ($(d1) + (1.5,0)$) {};
\draw (d1) -- +(-1,-0.5);
\draw (d1) -- +(-1,-0);
\draw (d1) -- +(-1,0.5);

\node[vertex] (d2) at (5,0) {};
\node[vertex] (d2p) at ($(d2) + (1.5,0)$) {};
 \node[vertex] (d3) at (10,0.5) {};
  \node[vertex] (d3p) at (10,-0.5) {};
    \node[vertex] (d3pp) at (11.5,-0) {};

\draw (d2) -- +(2,0);
\draw (d2) -- +(-1,-0.5);
\draw (d2) -- +(-1,0.5);
\draw (d2p) -- +(1,0);
 \draw (d3) -- (d3pp);
 \draw (d3p) -- (d3pp);
 \draw (d3) -- +(-1,0);
 \draw (d3pp) -- +(1,0.5);
 \draw (d3pp) -- +(1,-0.5);

\end{tikzpicture}
\caption{An Overview of Connected Contributions. The vertex represents a moduli space integral \cite{OP2,Troost:2026pmx}. It has a genus and a number of insertions that determine the class to integrate over the compactified moduli space. At the left, a degree zero purely left correlator. It can have a non-zero genus attached to the left vertex. In the middle, we have a degree one connected contribution. On the right, we sketched a degree $(1,1)$ connected contribution to a $(1,2)$-point function. }
\end{figure}
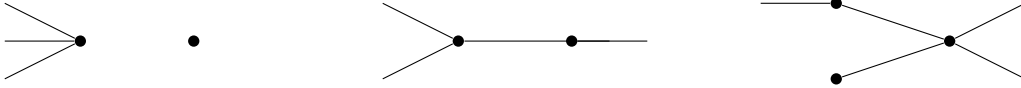
\noindent 
\subsection{The Connecting of Sheets at the Second Order}
At the next order in the $1/t$ expansion, which is order $O(t^{-4})$, we will  have non-trivial covers (of degree two) contributing to the perturbation theory. This allows for the interaction ${\cal F}_2$ to connect two macroscopic world sheets (instead of microscopic ones, which are mapped to a point).   At order $1/t^4$ in the expansion, we have the genus and degree combinations $(g,d)=(2,0)$, $(1,1)$ and $(0,2)$. The case $(g,d)=(2,0)$ are the familiar Hodge-dressed point theories. The $(g,d)=(1,1)$ interactions are the genus one corrections to the single cover world sheet instanton bridge we discussed previously. The genus zero, degree two interactions are qualitatively new. 
We will analyze the dependence of low-point functions as we vary $1/t$ from $0$ to $\infty$ and compare the contributions of disconnected covers and connected covers at degrees $d \ge 2$ in the next section.

\section{The Correlators and Connectedness}
\label{Correlators}
 We compute observables to confirm the intuition we started building about their behavior as we vary the coupling $g_s/t$. 
 We concentrate on zero, one, and two point functions at degrees zero, one, and two.  We compute these observables in the operator language \cite{OP2}. This formalism contains both connected and disconnected contributions, which will muddy the waters a little.

\subsection{The Zero Point Functions}
We exhibit the disconnected zero point function first. We have:
\begin{equation}
G_d(g_s) = \frac{g_s^{-2d}}{d!}  \, .
\end{equation}
The total genus of the amplitude equals $g=1-d$. We identify the contribution as $d$ disconnected degree one covers.

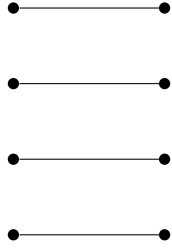
\begin{figure}[H]

\begin{tikzpicture}[
  scale=1,
  vertex/.style={circle, fill=black, inner sep=1.5pt}
]


\node[vertex] (d1) at (0,0) {};
\node[vertex] (d1p) at ($(d1) + (2,0)$) {};
\node[vertex] (d2) at (0,-1) {};
\node[vertex] (d2p) at ($(d2) + (2,0)$) {};
\node[vertex] (d3) at (0,-2) {};
\node[vertex] (d3p) at ($(d3) + (2,0)$) {};
\node[vertex] (d4) at (0,-3) {};
\node[vertex] (d4p) at ($(d4) + (2,0)$) {};

\draw (d1) -- +(2,0);
\draw (d2) -- +(2,0);
\draw (d3) -- +(2,0);
\draw (d4) -- +(2,0);

\end{tikzpicture}
\caption{The degree one disconnected covers of the sphere.}
\end{figure}
\noindent
The interaction term plays no role since it only modifies the operators, of which there are none. The equivariant parameter likewise does not enter since there is no non-trivial (equivariant) cohomology insertion. The grand canonical result for the zero point function equals \cite{OP2}:
\begin{equation}
\sum_{d=0}^\infty q^d G_d(g_s) = e^{\frac{q}{g_s^2}} \, .
\end{equation}
These degree $d$ zero point functions contribute to any disconnected $(n,m)$ point functions.

\subsection{The One Point Functions}
To compute the one-point functions, we recall the oscillator expression \cite{OP2}:
\begin{equation}
G_d(z,w,g_s) = g_s^{-2d-n-m} \langle A(tz_i,g_sz_i) e^{\alpha_1} P_d e^{\alpha_{-1}} A(-tw_j,g_sw_j)^\ast \rangle \, ,
\end{equation}
where $P_d$ is the operator that projects onto degree $d$ sheets \cite{OP2}.
Suppose we have only insertions of the first type. Then we must evaluate:
\begin{align}
G_d(z,g_s) &= \frac{1}{d!} g_s^{-2d-n} \langle \prod A(tz_i,g_sz_i) e^{\alpha_1} (\alpha_{-1})^d \rangle .
\end{align}
Because of the exponential factor, we have a staircase of contributions with $\prod A$ contributing from effective degree $0$ all the way up to effective degree $d$ (depending on the term we focus on in the expansion of the exponential). In particular, 
the (disconnected) Gromov-Witten one-point generating function at degree $d$ equals
\begin{equation}
G_d(z,g_s) = \frac{g_s^{-2d-1}}{d!}  \langle A(tz,g_sz) e^{\alpha_1} (\alpha_{-1})^d \rangle
\, .
\end{equation}
This is determined by the vacuum expectation values:
\begin{equation}
V_1 = \langle {\cal E}_i(g_sz) e^{\alpha_1} (\alpha_{-1})^d \rangle
\end{equation}
which gets contributions from terms in $e^{\alpha_1} = 1 + \alpha_1+\alpha_1^2/2 + \dots$ up to the term  $\alpha_1^d/d!$ and therefore from the operators ${\cal E}_{0 \le i \le d}$.
Consider the vacuum expectation value $V_1$ and pick the relevant term in the exponential:
\begin{align}
V_1&= \langle {\cal E}_{i}(g_sz) \frac{\alpha_1^{d-i}}{(d-i)!}(\alpha_{-1})^d \rangle =\langle {\cal E}_{i}(g_sz) \frac{d!}{i! (d-i)!}  (\alpha_{-1})^{i} \rangle
=\binom{d}{i} \langle {\cal E}_{i}(g_sz)  (\alpha_{-1})^{i} \rangle \, .
\end{align}
We moved the $\alpha_1$ oscillators to the right and counted contractions. 
Next, we use the commutation relation between the operators $\alpha_{-1}$ and ${\cal E}_k$ 
\begin{equation}
[{\cal E}_k(g_sz),\alpha_{-1}] = \zeta(g_sz) {\cal E}_{k-1}(g_sz) \, ,
\end{equation}
and find:
\begin{align}
V_1 &= \binom{d}{i} \zeta(g_sz)^{i} \langle {\cal E}_0 (g_sz) \rangle
\nonumber \\
&=  \binom{d}{i} \zeta(g_sz)^{i-1}
\, .
\end{align}
We plug this result back into the one-point function:
\begin{align}
G_d(z,g_s) &=\frac{g_s^{-2d-1}}{d!} S(g_sz)^{tz} \sum_{i=0}^d \binom{d}{i} \frac{\zeta(g_sz)^{2i-1}}{(tz+1)_{i}} \, .  \label{OnePointFunctionSummation}
\end{align}
This sum is a confluent hypergeometric function:
\begin{align}
G_d(z,g_s) 
&=\frac{g_s^{-2d-1}}{d!} S(g_sz)^{tz} 
\frac{1}{\zeta(g_sz)} \, \, {}_1F_1(-d,tz+1;-\zeta(g_sz)^2) \, . 
\end{align}
We analyze the result in the summation form (\ref{OnePointFunctionSummation}) at fixed degree $d$. 
At degree $d=0$, the result is:
\begin{equation}
G_0(z,g_s) = g_s^{-1} \frac{S(g_s z)^{tz}}{\zeta(g_s z)}
\, .
\end{equation}
At $t=0$, we find all the non-equivariant one-point functions given by the expectation value $1/\zeta(g_s z)$ of the operator ${\cal E}_0(g_s z)$. In the equivariant theory, the one-point functions are modified by the extra function $S(g_s z)^{tz}$. The extra factor  captures universal equivariant one-point functions that dress each operator separately.  
Next, let us analyze each term in the sum (\ref{OnePointFunctionSummation}). 
The term with $i=0$ equals the product of the zero point function discussed previously with the degree zero one-point function.  All terms with $i \ge 1$ are terms that have contracted the operator ${\cal E}_i$ with at least one sheet created by the $\alpha_{-1}$ operators.  Each of these terms will allow for any $tz$ power.\footnote{It may be useful to recall at this point that we must always expand in the variables $z_i$ first and only then analyze the large $t$ behavior.} However, note that all these terms come with a minimal power of $ g_s z$ determined by the power $2i-1$ of the $\zeta$ function.  Thus, the generic dominant term corresponds to the term with $i=1$ where we create a stable one-point function by attaching the operator to a single cover and otherwise complete the degree with extra disconnected covers. As we connect more and more covers, we augment $i$ and the terms are then subdominant at large $t$.

In the non-equivariant limit where $t=0$, on the other hand, we have that all terms are of the same order. The connected covers will contribute, as do the disconnected covers. The change in behavior is gradual. There is no phase transition between these two regimes -- indeed there are only a finite number of regular terms. 
If we have that the degree is larger or equal than two, we can compare the first two generic terms. We see that generically, when we compare the $i=1$ and $i=2$ contributions, we will have a $g_s^2/t^2$ correction when we compare the second type of term to the first. This is the price we pay to use two interaction terms $g_s/t \, {\cal F}_2$ to connect two sheets. 

\subsection{The Two-Point Functions}
The one-sided two-point functions exhibit the same recombination phenomena as we increase the $g_s/t$ coupling.   Let us demonstrate this explicitly.
To evaluate the generating function of two-point functions, we need the vacuum expectation values:
\begin{equation}
V_2=\langle {\cal E}_{l}(g_sz_1) {\cal E}_{-l+i}(g_sz_2) \alpha_{-1}^i \rangle
\, .
\end{equation}
For the value to be non-zero, we must have that $l \ge 0$. We can again move the $\alpha_{-1}$ oscillators to the left. We are then left with evaluating a ${\cal E}$ two-point function which is straightforward \cite{OP1}.  
%
We compute:
\begin{align}
V_2 &= \sum_{m=0}^i \binom{i}{m} \zeta(g_s z_1)^{m} \zeta(g_s z_2)^{i-m} \langle {\cal E}_{l-m}(g_s z_1) {\cal E}_{m-l} (g_s z_2) \rangle
 \\
 &= \delta_{l>m} \sum_{m=0}^{i} \binom{i}{m} \zeta(g_s z_1)^{m} \zeta(g_s z_2)^{i-m} \frac{\zeta((l-m)g_s(z_1+z_2))}{\zeta(g_s(z_1+z_2))} 
 \nonumber \\
 & + \delta_{l,m} \sum_{m=0}^{i} \binom{i}{m} \zeta(g_s z_1)^{m-1} \zeta(g_s z_2)^{i-m-1} \, .
 \nonumber
\end{align}
We find the two-point function:
\begin{align}
G_d(z_1,z_2,g_s,t) &= \frac{g_s^{-2d-2}}{d!}
\langle A(tz_1,g_s z_1) A(tz_2,g_s z_2) e^{\alpha_1} (\alpha_{-1})^d \rangle
\nonumber \\
&= \frac{g_s^{-2d-2}}{d!}\sum_{i=0}^d  \binom{d}{i}
\langle A(tz_1,g_s z_1) A(tz_2,g_s z_2)  \alpha_{-1}^i \rangle
\nonumber \\
&= \frac{g_s^{-2d-2}}{d!}S(g_s z_1)^{t z_1} S(g_s z_2)^{t z_2} \sum_{i=0}^d \binom{d}{i}
\sum_{m=0}^{i} \binom{i}{m}\nonumber \\
& \Big( \sum_{l > m} 
\frac{\zeta(g_s z_1)^{l+m} }{(tz_1+1)_{l}} \frac{\zeta(g_s z_2)^{-l-m+2i}}{(t z_2+1)_{i-l}} 
  \frac{\zeta((l-m)g_s(z_1+z_2))}{\zeta(g_s(z_1+z_2))} 
 \nonumber \\
 & +
 \frac{\zeta(g_s z_1)^{2m-1} }{(tz_1+1)_{m}} \frac{\zeta(g_s z_2)^{-2m+2i-1}}{(t z_2+1)_{i-m}} \Big )
 \, .
\end{align}
We surmise again that we want the power of $z_1$ plus the power of $z_2$ coming from the $\zeta$ functions to be minimal in the large $t$ limit. The last term, when it is non-zero, will have the minimal power of $(z_1,z_2)$, for the value $i=0$. Therefore, the minimal value $i=0$  again instructs us to look at the point theory and in particular at the disconnected contribution when non-zero. Higher powers of $(z_1,z_2)$ obtained by further contractions between insertions and edges will  lead to subdominant contributions. The generic behavior is that when we increase the counting parameter $i$, i.e. the number of sheets we connect to the operators, the contribution becomes more and more subdominant in the large $t$ limit.   The rest of the discussion is as before - the recombination at higher orders in $1/t$ is a universal phenomenon. We also recognize the combinatorial factors that pick $d-i$ out of $d$ covers to form independent disconnected world sheets as well as further choices as to which remaining world sheets shall be decorated with which markings.
%

\subsection{The Relative One-Point Functions}
In this subsection, we study the relative one point function using the results of \cite{OP2,OP3}. The relative one point function  measures the weight of each partition $\nu$ inside a state generated by one vertex operator. 
The relative one-point function is:
\begin{equation}
G(z|\nu;g_s,t) = \sum_k z^{k+1} \langle \tau_k[0]
|\nu \rangle = \langle A(zt,g_st) e^{\alpha_1} | \nu \rangle \, .
\end{equation}
We compute the one-point function for a generic profile $\nu=\{ \nu_i \}$. For a  profile that contains a power $m_1$ of $\alpha_{-1}$ oscillators,  we can insert a number $n_1$ of $\alpha_1$ oscillators that is smaller than or equal to the power $m_1$. We find:
\begin{align}
G(z|\nu;g_s,t) &= \sum_{n_1=0}^{m_1} \frac{1}{n_1!} \langle A(zt,g_st) | {\nu}'(n_1)\rangle \, ,
\end{align}
where ${\nu}'$ is the partition $\nu$ where we eliminated $n_1$ ones.  To compute the prefactor, it is necessary to recall the automorphism factor $z(\nu)$ which normalizes the state $\nu$:
\begin{equation}
|\nu \rangle = \frac{1}{z(\nu)} \prod_i \alpha_{-\nu_i} | 0 \rangle \, .
\end{equation}
 Only the term 
\begin{equation}
V_4 = \langle {\cal E}_{|{\nu}'|} (g_sz) | {\nu}' \rangle
\, 
\end{equation}
in the operator $A$
contributes.
We take all of the oscillators $\alpha_{-\nu_i'}$ in the state on the right and commute them through to the left. We find:
\begin{equation}
V_4 = \frac{1}{z({\nu}')}  \left( \prod_{i'} \zeta(\nu_i' g_s z) \right) \frac{1}{\zeta(g_s z)} \, .
\end{equation}
We plug this into the operator formula for the one-point function:
\begin{align}
G(z|\nu;g_s,t) &=\frac{1}{z(\nu)} \sum_{n_1=0}^{m_1} \binom{m_1}{n_1} S(g_s z)^{t z} 
\frac{\zeta(g_s z)^{|\nu'|-1}}{(tz+1)_{|\nu'|}} \left( \prod_{i'} \zeta(\nu_i' g_s z) \right) 
\, . \label{OnePointExplicit}
\end{align}
In the large $t$ limit, we again find the dominant term by minimizing the powers of $g_s z$ in order to maximize the power of $t z$ for a given correlation function. That implies that we must minimize $|\nu'|$ which is done by maximizing $n_1$. Therefore, we take out all the $\alpha_{-1}$ oscillators from the initial state $|\nu \rangle$ by maximizing $n_1=m_1$. Thus, if we fix the $\nu$ profile, the amplitude is again dominated by the diagrams with disconnected covers of degree one at large $t$. The pattern is universal for both relative and absolute correlators. 

\section{The Near Non-Equivariant Theory}
\label{NearNonEquivariant}
In a final exploration of the equivariant $\mathbb{P}^1$ theory, we study the bulk theory perturbed around the non-equivariant point. We use the bulk oscillator description of correlators and perform their expansion 
at small $t$. We start once more from the operator insertions:
\begin{equation}
A(tz,g_sz) = S(g_sz)^{tz} \sum_{k \in \mathbb{Z}} 
\frac{\zeta(g_sz)^k}{(tz+1)_k} {\cal E}_k(g_sz) \, .
\label{OperatorsA}
\end{equation}
We know that at the non-equivariant point $t=0$, the entire operator collapses to \cite{OP2}:
\begin{equation}
A(0,g_sz) =  \sum_{k \in \mathbb{N}} 
\frac{\zeta(g_sz)^k}{k!} {\cal E}_k(g_sz)= e^{\alpha_1} {\cal E}_0(g_sz) e^{-\alpha_{1}}\, .
\label{Aat0}
\end{equation}
From this operator equality, we obtain the expression for the non-equivariant $n$-point function at degree $d$ \cite{OP1,OP2}:
\begin{equation}
G_d(z) = \frac{1}{(d!)^2} \langle \alpha_1^d \prod {\cal E}_0(g_sz_i) \alpha_{-1}^d \rangle \, .
\end{equation}
We want to compute the  order $O(t^2/g_s^2)$ corrections to this result. Note that these are corrections captured by lower genus Riemann surfaces. Thus, from a string theory perspective, we are expanding around a strong coupling point. Recall that it is at the strong coupling point that the Gromov-Witten theory is equivalent to a grand canonical Hurwitz theory \cite{OP1}. 
When we analyze the corrections, we express them in terms of the operators ${\cal E}_0(g_s z)$ that code the volume descendants at the strong coupling point, and the operators $\alpha_{\pm 1}$ that create single disconnected sheets. 
The corrections will arise from modifications of the operators $A(0,g_sz_i)$. 
\subsection{The Prefactor}
We have a first type of corrections that renormalize all operators. These arise from the prefactor:
\begin{align}
S(g_sz)^{tz} & = \exp (t z \log S(g_sz)) \approx 1+ tz \log S(g_sz) + \frac{t^2z^2}{2} \log^2S(g_sz) + O(t^3) 
\nonumber \\
&= 1+\frac{1}{24} t g_s^2 z^3-\frac{z^5 \left(t g_s^4\right)}{2880}+\frac{t^2 g_s^4 z^6}{1152}+\frac{t g_s^6 z^7}{181440}
\nonumber \\
 & -\frac{z^8 \left(t^2 g_s^6\right)}{69120}+\frac{z^9 \left(350 t^3 g_s^6-3 t g_s^8\right)}{29030400}+O\left(z^{10}\right) \, .
\end{align}
The prefactor is easily expanded and if desired  can be soaked up in a universal redefinition of the operators. It codes  tadpoles. For example the $z^3$ term corresponds to a $\tau_1$ equivariant one-point function. 
\subsection{The Expanded Operators}
We turn to the more interesting modifications of the operator $A$ when expanded around $t=0$. 
At zeroth order in the parameter $t$, only the operators ${\cal E}_k$ with $k \ge 0$ have a non-zero coefficient in the operator $A$ -- see equations (\ref{OperatorsA}) and (\ref{Aat0}). Therefore,  the non-equivariant correlators simplify to only $k=0$ insertions (after commutation with $e^{\pm \alpha_1}$). At the first order in $tz$ already, all terms ${\cal E}_k$ have a non-zero coefficient, for any $k \in \mathbb{Z}$. Then, all terms matter. 
The perturbation theory is rendered complicated because to the ${\cal E}_0$ terms at zeroth order, we add an infinite number of extra terms and contractions, even at first order. We assuage these complications by immediately concentrating on absolute and relative one-point functions. In that case, we can restrict the analysis to terms with operators ${\cal E}_k$ with $k \ge 0$. 
In that range, we write:
\begin{equation}
\zeta(g_s z)^k {\cal E}_k(g_s z) =
[\alpha_1,[\dots,[\alpha_1,{\cal E}_0 (g_s z)]]]]
\, ,
\end{equation}
where we have $k$ commutators. 
We also use the expansion of the Pochhammer symbol for
 $k \ge 0$:
\begin{equation}
\frac{1}{(tz+1)_k} = \frac{1}{k!} \left( 1- H_k \, tz + \frac{1}{2 } (H^2_k+H^{(2)}_k) (tz)^2  + \dots \right) \, .
\end{equation}
The harmonic numbers $H_k^{(n)}$  are defined as
\begin{equation}
H_k^{(n)} = \sum_{l=1}^k \frac{1}{l^n}
\, ,
\end{equation}
with the notation $H_k=H_k^{(1)}$.  
We already know how to represent the first term in the $k \ge 0$ summation as an exponential conjugation -- see equation (\ref{Aat0}). We can perform a similar rewriting for the next terms using the integral representations:
\begin{equation}
H_k = \int_0^1 \frac{1-s^k}{1-s} \, ds
\end{equation}
as well as:
\begin{equation}
H_k^{(2)} = \int_0^1 \frac{1-s^k}{s-1}  \log s \, ds \, .
\end{equation}
We plug all these results into the (renormalized and truncated) expression for the operator:
\begin{align}
\tilde{A}^{\ge 0}
&= \sum_{k \ge 0} \frac{\zeta(g_s z)^k}{(tz+1)_k} {\cal E}_k(g_s z)
%
\nonumber \\
&=
\sum_{k \ge 0}^\infty \left( 1-\int_0^1 \frac{1-s^k}{1-s} tz +\frac{1}{2} \left(
(\int_0^1 ds\frac{1-s^k}{1-s})^2 + \int_0^1 \frac{1-s^k}{s-1} \log s \, ds  \right) (tz)^2 +\dots \right) 
\nonumber \\
& \qquad  \frac{1}{k!} [\alpha_1,[\dots,[\alpha_1,{\cal E}_0(g_s z)]]]
\nonumber \\
&=  e^{\alpha_1} {\cal E}_0(g_s z) e^{-\alpha_1} -(\int_0^1 ds
\frac{1}{1-s} ( e^{ \alpha_1} {\cal E}_0 e^{-\alpha_1}- e^{s \alpha_1} {\cal E}_0 e^{-s\alpha_1} ))tz
 + O(t^2 z^2) \, .
\end{align}
The order $t^2 z^2$ term can be similarly represented as a conjugation of the operator ${\cal E}_0$. The main difference will be the appearance of a double integral arising from the square of the harmonic number. The integral representation is geared towards combining the exponentials of $\pm \alpha_1$ factors in the correlator, as we did at zeroth order. 

Given this representation of the operator $A$, we can provide a perturbative series for the relative one-point functions. To first order in perturbation theory, one finds after a change of variables $\tilde{s}=1-s$:
\begin{align}
G(z) &= \langle A e^{\alpha_1} | \nu \rangle
\nonumber \\
&= S(g_s z)^{t z} \langle e^{\alpha_1} \left( {\cal E}_0(g_s z) - \int_0^1 \frac{ds}{s} ({\cal E}_0-e^{-s \alpha_1} {\cal E}_0 e^{s \alpha_1}) tz + \dots \right) | \nu \rangle \, . \label{PerturbativeAtZero}
\end{align}
Thus, we manifestly find the leading unperturbed result as well as the first order correction which we managed to write as a prefactor dressing plus a perturbing operator which is made up of the operators ${\cal E}_0$ capturing excitations in the unperturbed theory conjugated by integrated exponentials of the operator $\alpha_1$. This pattern will extend to higher orders, with multiple integrals appearing. Harmonic number coefficients are guaranteed to appear in the perturbation theory. Higher $n$-point functions on the other hand, will require new ingredients. 
For the relative one-point function, we note that
the first interaction term can peel off $\alpha_{-1}$ oscillators from the $|\nu\rangle$ relative state and then contribute. This is a phenomenon that we can confirm with the exact expression for the relative one-point function which is readily available in this simple case -- see (\ref{OnePointExplicit}) :
\begin{align}
G(z|\nu;g_s,t) &=\frac{1}{z(\nu)} \sum_{n_1=0}^{m_1} \binom{m_1}{n_1} S(g_s z)^{t z} 
\frac{\zeta(g_s z)^{|\nu'|-1}}{(tz+1)_{|\nu'|}} \left( \prod_{i'} \zeta(\nu_i' g_s z) \right)  \, .
\end{align}
The first correction to the non-equivariant result depends on the positive inverted Pochhammer symbol as well as the prefactor.
The subleading terms can arise from all types of cycles, that is to say, after peeling off $n_1$ out of $m_1$ $1$ contributions to the partition $\nu$. 
These are the leading contributions  captured by the perturbative expression (\ref{PerturbativeAtZero}). 

In conclusion, the naive approach to perturbation theory shows the appearance of the harmonic numbers in perturbation theory and perturbing operators, which are integrated conjugations of the non-equivariant operators ${\cal E}_0$ by exponentials of $\alpha_1$. The strong coupling perturbation theory will rapidly become unwieldy.
It may be useful to attack this problem under another angle. Since the interaction term is of the form 
$\exp(g_s/t {\cal F}_2)$ one may want to formulate a saddle point approximation at small equivariant parameter $t$. 

\subsection{Summary and Remarks}
Let us take stock of the analyses performed in the last four sections. We have argued that at infinite equivariant parameter, the Gromov-Witten theory on $\mathbb{P}^1$ reduces to two pure topological gravity theories associated to the two fixed points. Genus corrections dress the theories with Hodge classes. Covers of the sphere appear in perturbation theory in $q/t^2$ and constitute a coupling between two pure theories of gravity. 

We find it interesting that we have a two-dimensional theory of gravity at hand which can be interpreted as a consistent coupling of two factorized models of pure gravity. Initially separate universes are connected by world sheet instantons, consistently with two-dimensional quantum gravity. The perturbation theory shows that disconnected covers of the sphere dominate at large $t$ while at $t$ of order $g_s$, all types of connected contributions will enter the amplitude. At strong coupling, near $t=0$, another exact dual appears in the form of the grand canonical Hurwitz theory. While one can organize perturbation theory at small $t$ in terms of the operators of the Hurwitz theory, the perturbation theory is intricate. It is nevertheless interesting, since it allows, in principle, to start from the Hurwitz dual and formulate a holographic duality for non-zero equivariant parameter. Clearly, a similar attempt to reinterpret the perturbation theory near large equivariant parameter $t$ in terms of the dual pair of Kontsevich models would also be worthwhile -- one would need to come to terms with formulating the world sheet instanton perturbation in the matrix model language. 

\section{The Equivariant Cohomology of the Hilbert Scheme}
\label{EquivariantHilbertScheme}

In this section, we move to another corner of the landscape sketched in section \ref{EquivariantLandscape} and study aspects of the  correspondence between Gromov-Witten theory on $\mathbb{P}^1 \times \mathbb{C}^2$ and the equivariant cohomology of the Hilbert scheme of points on the complex two-plane $\mathbb{C}^2$. We concentrate on the boundary theory and review the construction of the equivariant ring and how to determine the structure constants. We then show how to think of the equivariant ring as a deformation of the ordinary cohomology ring on the one hand, and on the other, discuss some remarkable properties of the equivariant structure constants. 

The ordinary cohomology of the Hilbert scheme of points is rather well-known in physics since it features, for instance, as a subspace of highly supersymmetric states \cite{Dijkgraaf:1996xw} in  two-dimensional conformal field theory or in the moduli space of instantons of supersymmetric gauge theories. The equivariant cohomology  also finds natural applications in physics whether through an equivariant deformation of the moduli space of instantons \cite{Nekrasov:2002qd} or through its relation to $\beta$ ensembles in matrix models and conformal field theory \cite{Okounkov,Morozov:2013rma}. In the following, we concentrate on the question of to what degree we can concretely compute the structure constants of the equivariant cohomology ring, and we review combinatorial conjectures and theorems about their  properties. Recent mathematical results provide considerable insight into the physical question of what the structure constants of the ring are, and the physical background we provide  further motivates some of the mathematical constructions.

\subsection{The Hilbert Scheme of Points}

The Hilbert scheme of points $\text{Hilb}^n(\mathbb{C}^2)=X^{[n]}$ on the complex two-plane can be defined as the set of ideals of the polynomial ring $\mathbb{C}[x,y]$ which are of dimension zero and length $n$. The space of such ideals is $2n$ dimensional and smooth \cite{Fogarty}. It therefore has a well-defined cohomology. The smooth manifold allows for a description of the de Rham model in terms of a supersymmetric quantum mechanics \cite{Witten:1982df}.
More concretely,  the calculation of the standard cohomology can be performed using a Morse function (see e.g. \cite{Morse,Witten:1982im,NakajimaLectures}) 
or a contraction (or Bialynicki-Birula  decomposition) of the space \cite{ES}. The ring structure on the cohomology is considerably more difficult to compute
\cite{LS}. We reviewed the ring structure as well as its applications, including the calculation of an infinite set of new structure constants in the ring of chiral primaries for the associated topological conformal field theory in \cite{Li:2020zwo,Ashok:2023mow} to which we refer for background on the present chapter as well. To describe the structure constants very concretely, it turned out to be useful to think of the ring in terms of symmetric functions \cite{LS} and their relation to combinatorics. In this section, we wish to extend these hands-on insights into the ordinary cohomology to the equivariant cohomology.\footnote{As described previously, one motivation is that these structure constants can be thought off as two parameter generalizations of the three-point functions arising in the non-equivariant $AdS_3/CFT_2$ correspondence.}  Recent results by combinatorialists are very relevant for the physical question of calculating the product of operators in the cohomology. 
  
\subsection{The Equivariant Cohomology}
\label{EquivariantCohomologyHilbertScheme}

The equivariant cohomology of the Hilbert scheme of points and its relation to symmetric polynomials was established in \cite{NakajimaJack,Vasserot}. In our description of the calculation, we were guided by the review  \cite{LQW}. There is also a good summary in \cite{NakajimaMoreLectures}. We refer to these references for more background. We will briefly recall how to move from the application of the equivariant localization theorem \cite{AB} to a concrete description of the structure constants of the ring. 

The calculation of the equivariant cohomology starts out with the choice of a 
  $G=U(1)$ group action given by two non-negative integers $\alpha,\beta$ on the two-dimensional complex plane parameterized by $(x,y)$: 
\begin{align}
s\cdot (x,y) &= (s^\alpha  x,s^{-\beta} y) \label{U(1)Action}
\end{align}
where $s \in U(1)$.\footnote{The action can be extended to a $U(1) \times U(1)$ group action if desired. We follow the mathematical literature and keep $\alpha$ and $\beta$ general for the time being.}
We distinguish two two-planes. The two-plane $\Sigma$ is given by the equation $y=0$, and the two-plane $\Sigma'$ corresponds to the submanifold $x=0$. 
Due to the transformation rule (\ref{U(1)Action}), the tangent space to the real two-plane $\Sigma$ carries a character of the $U(1)$ action which is $\theta^{-\alpha}$ where $\theta$ is the representation of the character with charge one. 
The only fixed point of the $U(1)$ action in the original $\mathbb{C}^2$ manifold is the origin $o$. 
The equivariant localization theorem \cite{AB} implies the equivalence of cohomology classes
\begin{align}
[\Sigma] &= - \frac{1}{\alpha t} [o] \nonumber \\
[\Sigma'] &= \frac{1}{\beta t} [o] \, ,
\end{align}
where the class $[\Sigma]$ is determined by the Gysin map associated to the inclusion $i$ in $\mathbb{C}^2$ (and we have generically $i : N \rightarrow M$ and $[N]=i_{!}(1_N)$). The denominator arises from the Euler class of the normal bundle and we have denoted the equivariant parameter by $t$. 
In the Hilbert scheme on the other hand, the torus fixed points are isolated and labelled by partitions $\lambda$ of $n$ corresponding to ideals generated by monomials \cite{NakajimaLectures}. To apply the localization theorem, we need  the $U(1)$ action on the tangent space to the Hilbert scheme of $X=\mathbb{C}^2$. The tangent space at the fixed point $\xi_\lambda$ associated to the partition $\lambda$ is isomorphic to \cite{ES,NakajimaLectures}:
\begin{equation}
T_{\xi_\lambda} (X^{[n]}) = \bigoplus_{s \in \lambda} 
\left( \theta^{\alpha (l(s) + 1) + \beta a(s)} \oplus
 \theta^{-\alpha l (s) - \beta(a(s)+1)} \right) \, ,
 \label{TangentSpaceAtFixedPoint}
\end{equation}
where by a standard abuse of notation $\lambda$ is the Young diagram of the partition $\lambda$, the sum is over the boxes of the diagram and $a,l$  denote the arm and leg of the box \cite{Macdonald} -- see also Appendix \ref{Symmetric}. 
As a consequence, the equivariant Euler class at a given fixed point is:
\begin{equation}
E_T (T_{\xi_\lambda} (X^{[n]}))= (-1)^n \tilde{c}_\lambda \tilde{c}_\lambda' t^{2n} \, ,
\end{equation}
where the constants $\tilde{c}_\lambda,\tilde{c}_\lambda'$ are defined as \cite{LQW}:
\begin{align}
\tilde{c}_\lambda(\alpha,\beta) &= \prod_{s \in \lambda} (\alpha (l(s) + 1) + \beta a(s)) \nonumber \\
\tilde{c}'_{\lambda}(\alpha,\beta) &= \prod_{s \in \lambda} (\alpha l (s) + \beta(a(s)+1))
\, .
\end{align}
To describe the equivariant ring, we concentrate on classes $[\lambda]$ in the cohomology $H_T^{2n}(X^{[n]})$ \cite{NakajimaJack,Vasserot} which are defined in terms of the classes associated to the fixed points $\xi_\lambda$:
\begin{equation}
[\lambda] = \frac{(-1)^n}{\tilde{c}_\lambda} t^{-n} [\xi_\lambda] \, .
\label{ClassNormalization}
\end{equation}
Because the Hilbert scheme $X^{[n]}=\text{Hilb}^n(\mathbb{C}^2)$ only has non-zero even Betti numbers and because $H^k(X^{[n]})=0$ for $k > 2n$, one can prove that  $H^{2k \ge 2n}_T = t^{k-n} \cup H^{2n}_T$, namely that every higher degree form starts out with an appropriate number of powers of the equivariant parameter $t$ of degree two. We can then define a ring structure $\ast$ on middle cohomology on the basis of the cup product $\cup$ as follows \cite{Vasserot}:
\begin{equation}
t^{n} \cup (A \ast B) = A \cup B \, 
\label{DefinitionProduct}
\end{equation}
for $A,B \in H^{2n}(X^{[n]})$. 
There is also a 
bilinear form on the middle cohomology:
 \begin{equation}
 \langle [\lambda], [\mu] \rangle = \delta_{\lambda,\mu} \frac{\tilde{c}'_\lambda}{\tilde{c}_\lambda} 
 \label{BilinearForm}
 \, .
 \end{equation}
 The bilinear overlap is straightforward to understand.
The Euler class in the localization theorem contributes factors $\tilde{c}_\lambda \tilde{c}'_\lambda$ and the normalization of the classes the factors $1/(\tilde{c}_\lambda \tilde{c}_\mu)$. Fixed point classes moreover only overlap if they are localized at the same fixed point. The result follows.  The bilinear form is non-degenerate.

To understand the ring structure  better,  we briefly summarize a large body of work on the structure of the equivariant cohomology of the Hilbert scheme of points that was established by analyzing the structure of the cup product with respect to multiplication by classes associated to subvarieties. Firstly, Heisenberg operators $p_i$ associated to the 
curve $\Sigma$ were defined \cite{NakajimaLectures,Grojnowski}.\footnote{Note that this treats one plane preferentially over another  -- we will see interesting consequences of this in the sequel. }
The Heisenberg algebra has a Planck constant defined by the overlap
$\langle [\Sigma],[\Sigma] \rangle = \beta/\alpha$ which follows from  the general formulas for the inner product (\ref{BilinearForm}) applied to the case $n=1$ and $\tilde{c}_{[1]}(\alpha,\beta)=\alpha,\tilde{c}'_{[1]}(\alpha,\beta)=\beta$.
If we define operators
\begin{align}
p_{-\lambda} &= 
\prod_{i \ge 1} p_{-i}^{m_i} \, ,
\end{align}
the Heisenberg algebra implies for the corresponding cohomology states \cite{NakajimaMoreLectures}:
\begin{align}
\langle p_{-\lambda} |0 \rangle, p_{-\mu} | 0 \rangle \rangle &= \delta_{\lambda,\mu} {z(\lambda)}  (\frac{\beta}{\alpha})^{l(\lambda)} \, .
\label{Overlap}
\end{align}
These cohomology classes correspond to power sum symmetric functions.\footnote{Again, see Appendix \ref{Symmetric} for a review of symmetric functions and their properties.} 
Moreover, intermediate geometric classes $[\Sigma_\lambda]$ can be defined, associated to components $\lambda$ of the symmetric product of $\Sigma$ such that they are isomorphically mapped  to the monomial symmetric functions $m_\lambda$.
One further has the triangular decomposition \cite{NakajimaJack,Vasserot}: 
\begin{equation}
[\Sigma_\lambda] = [\lambda]+ \sum_{\mu < \lambda} c_{\lambda \mu} [\mu] \, ,
\label{Triangular}
\end{equation}
where we used the dominance partial ordering $<$. All these facts together (and in particular the overlap (\ref{Overlap}) and the triangular structure (\ref{Triangular})), combined with the theory of Jack symmetric functions \cite{Macdonald}, implies that we have the isomorphism $\phi$ between equivariant cohomology classes and symmetric functions \cite{NakajimaJack,Vasserot,LQW,NakajimaMoreLectures}:
\begin{equation}
\phi(p_{-\lambda} |0 \rangle) = p_\lambda \, , \qquad \phi([\Sigma_\lambda]) = m_\lambda \, ,
\qquad \phi([\lambda]) = 
P_\lambda^{(\beta/\alpha)}
\, , 
\label{Isomorphism}
\end{equation}
where $p_\lambda$ are the power sum polynomials and $P_\lambda^{(\beta/\alpha)}$ are the Jack symmetric functions. 
Finally, the product of classes is calculated through the localization theorem \cite{AB,LQW}:
\begin{equation}
[\xi_\lambda] \cup [\xi_\mu] = (-1)^n \delta_{\lambda,\mu} \tilde{c}_\lambda(\alpha,\beta) \tilde{c}'_\lambda(\alpha,\beta) t^{2n} [\xi_\lambda] \, ,
\end{equation}
where the delta-function codes that there is only a non-trivial overlap between classes at the same fixed point and all the prefactors arise from the Euler class in the localization formula \cite{LQW}. Using the definition  of the product (\ref{DefinitionProduct}) and the normalization (\ref{ClassNormalization}) then implies that the classes $[\lambda]$ multiply as:
\begin{equation}
\frac{[ \lambda]}{\tilde{c}_\lambda'} \ast \frac{[ \mu]}{\tilde{c}_\mu'} = \delta_{\lambda,\mu} \frac{[ \lambda]}{\tilde{c}'_\lambda} \, .
\end{equation}
Therefore, we define a product $\circ$ on Jack polynomials such that it coincides with the ring product $\ast$ under the isomorphism $\phi$ in (\ref{Isomorphism}):
\begin{align}
\frac{P_\lambda^{(\beta/\alpha)}}{\tilde{c}_\lambda'}
\circ \frac{P_\mu^{(\beta/\alpha)}}{\tilde{c}_\mu'}
&= \delta_{\lambda,\mu} \frac{P_\lambda^{(\beta/\alpha)}}{\tilde{c}_\lambda'} \, . \label{ProductOfJackFunctions}
\end{align}
This defines an interesting product $\circ$ on the Jack polynomials.  The cohomological ring structure on Jack polynomials makes them idempotents -- a direct consequence of the fact that they are associated to fixed points.
The main conclusion from this subsection is that the equivariant cohomology ring is isomorphic to the space of Jack polynomials with the multiplication $\circ$. 

In the following, we sometimes use an alternative notation for the parameters $-\alpha t=t_1$ and $\beta t = t_2$ and think of $t_{1,2}$ as independent continuous numbers.
We shall also use the notation  ${\alpha}_J = \beta/\alpha= -t_2/t_1$ for the $\alpha_J$ parameter of the Jack polynomials $P^{(\alpha_J)}_\lambda$.\footnote{Moreover, we will drop the index $J$ when there can be no confusion between the Jack parameter and the equivariant parameter $\alpha$. }

\subsection{The Ordinary Cohomology Revisited}
\label{OrdinaryCohomologyHilbertScheme}
The  product $\circ$ on the Jack polynomials $P_\lambda^{(\beta/\alpha)}$ deserves  elucidation. In this subsection, we provide insight into this product by proving that it gives rise to the ring structure of the ordinary cohomology ring in the limit where we send the equivariant parameters $t_i$ to zero.
To understand the  limit theory, we
recall that the ordinary cohomology ring structure constants had an efficient incarnation in terms of top connection coefficients of class functions in the symmetric group algebra \cite{LS}. 
Thus, we review basic facts about class functions and their relation to symmetric polynomials. See \cite{Li:2020zwo} for considerably more physics background.

\subsubsection{Symmetric Group Theory}
We view the ordinary cohomology ring as isomorphic to the center of the group algebra of $S_n$ and define a basis of class functions $C_\mu$ which are sums over the elements of conjugacy classes $C^\mu$ with coefficient one:
\begin{align}
C_\mu &= \sum_{\sigma \in C^\mu}  \sigma \, .
\end{align}
We introduce the quantities  $\chi^\lambda_\sigma=\chi^\lambda(\sigma)$ which are the  characters $\chi^\lambda$ of the representation $\lambda$ evaluated at a group element $\sigma \in S_n$. 
The idempotent linear combinations $e_\mu$ of the class functions $C_\mu$ are  \cite{HSS}:
\begin{equation}
e_\mu = \frac{d_\mu}{|S_n|} \sum_\lambda \bar{\chi}^\mu_\lambda C_\lambda 
\end{equation}
where we consider $\bar{\chi}$ to be the character $\chi$ evaluated at the inverse group element (or the conjugate character) and $d_\mu$ is the dimension of the representation $\mu$. 
The variable $\lambda$ runs over the conjugacy classes of the symmetric group $S_n$ or equivalently, over the partitions of $n$.  The fact that these are idempotents, namely that they satisfy
\begin{equation}
e_\mu e_\nu = \delta_{\mu,\nu} e_\nu
\, ,
\end{equation}
is non-trivial and uses the fact that the functions $C_\mu$ are central in the group algebra.\footnote{For a detailed proof, see chapter XV of \cite{Burnside}.}
Using the orthogonality relations for characters, we can express the class sums in terms of the idempotents:
\begin{equation}
C_\mu = |C_\mu| \sum_\lambda \frac{\chi_\mu^{\lambda}}{d_\lambda} e_\lambda \, ,
\end{equation}
where $|C_\mu|$ is the number of group elements in the class $C^\mu$.
Moreover, the  product of conjugacy class sums satisfies:
\begin{align}
C_\mu C_\nu &= |C_\mu| |C_\nu| \sum_\lambda \frac{\chi^{\lambda}_\mu \chi_\nu^{\lambda}}{d_\lambda^2} e_\lambda
\nonumber \\
 &=  \frac{|C_\mu| |C_\nu|}{|G|} \sum_\lambda \frac{\chi^{\lambda}_\mu \chi_\nu^{\lambda}}{d_\lambda} \sum_\rho \bar{\chi}_\rho^{\lambda} C_\rho
 \nonumber \\
 &= \sum_\rho \left( \frac{|C_\mu| |C_\nu|}{|G|} \sum_\lambda \frac{\chi^{\lambda}_\mu \chi_\nu^{\lambda} \bar{\chi}_\rho^{\lambda}}{d_\lambda}  \right) C_\rho \, .
\end{align}
We define structure constants $a$ such that:
\begin{equation}
C_\mu C_\nu = {a_{\mu \nu}}^\rho C_\rho
\, .
\end{equation}
Let $h_\lambda$ be the product of hook lengths of the boxes in the diagram $\lambda$ \cite{Macdonald}. Then the dimension of the representation $\lambda$ is $d_\lambda=n!/h_\lambda$. The size of the centralizer of an element in the conjugacy class $\lambda$ is $z_\lambda = n!/|C_\lambda|$. Note moreover that the  characters of $S_n$ are real. We then find the structure constants:
\begin{equation}
{a_{\mu \nu}}^\rho = \frac{1}{z_\mu z_\nu} \sum_\lambda h_\lambda \chi^\lambda_\mu \chi^\lambda_\nu \chi^\lambda_\rho \, .
\label{ClassConnectionCoefficients}
\end{equation}
The ring product in the ordinary cohomology is the convolution product restricted to those terms that  preserve the degree $d(\lambda)=n-l(\lambda)$ where $l(\lambda)$ is the length of the partition $\lambda$ \cite{LS}. The cohomology product therefore arises from the convolution product by filtration with respect to this degree. We plan to understand how the formula for the structure constants as well as the filtration arise  from taking a limit on the equivariant cohomology structure constants. 

\subsubsection{Schur Polynomials and Connection Coefficients}
It is possible to establish an isomorphism between the conjugacy class sums and the power sum symmetric polynomials \cite{LS}.  The product $\ast_{conv}$ of the power sum polynomials is determined by the ring isomorphism  and therefore has structure constants equal to the connection coefficients:
\begin{equation}
\frac{p_\lambda}{z_\lambda} \ast_{conv} \frac{p_\mu}{z_\mu} = {a_{\lambda \mu}}^\nu \frac{p_\nu}{z_\nu} \, .
\end{equation}
To make the link between the equivariant cohomology and the ordinary cohomology, we first show that the $\circ$ product (\ref{ProductOfJackFunctions})  on Jack functions reduces to the convolution product $\ast_{conv}$ up to an overall factor, in the case where we consider an equal and opposite action of the $U(1)$ on the two-planes $\Sigma$ and $\Sigma'$, namely when we restrict to $\alpha=\beta$.  
Indeed, we note that the  Jack functions  $P_\lambda^{(\beta/\alpha)}$ reduce to the Schur functions $P_\lambda^{(1)}=s_\lambda$ \cite{Macdonald} on the subspace  $\alpha=\beta$. Under this assumption, we find the $\circ$ product on Schur functions:
\begin{equation}
\frac{s_\lambda}{\alpha^n h_\lambda}
\circ \frac{s_\mu}{\alpha^n h_\mu}
= \delta_{\lambda,\mu} \frac{s_\lambda}{\alpha^n h_\lambda}  \, ,
\end{equation}
where $h_\lambda$ is the product of hook lengths \cite{Macdonald} of the boxes in the Young diagram of $\lambda$, since $\tilde{c}_\lambda(\alpha,\alpha)=\alpha^n h_\lambda$. 
Moreover, we have the change of basis from Schur to power sum symmetric functions, given by the characters \cite{Macdonald}:
\begin{align}
s_\lambda &= 
\frac{1}{z_\mu} \chi^\lambda_\mu \, p_\mu
\, .
\end{align}
We will regularly leave out the summation symbols from now on.
We also have the inverse relation:
\begin{equation}
p_\rho = \chi^\lambda_{\rho} s_\lambda \, .
\end{equation}
This allows us to compute the product $\circ$ of power sum symmetric functions:
\begin{align}
\frac{p_\lambda}{z_\lambda} \circ \frac{p_\mu}{z_\mu} &=
\frac{\chi^{\lambda_1}_\lambda}{z_\lambda} s_{\lambda_1}
\circ \frac{\chi^{\lambda_2}_\mu}{z_\mu} s_{\lambda_2}
= \alpha^n \frac{\chi^{\lambda_1}_\lambda}{z_\lambda} 
 \frac{\chi^{\lambda_1}_\mu}{z_\mu} h_{\lambda_1} s_{\lambda_1} 
\nonumber \\
&= \alpha^n \frac{\chi^{\lambda_1}_\lambda}{z_\lambda} 
 \frac{\chi^{\lambda_1}_\mu}{z_\mu} h_{\lambda_1} 
 \frac{\chi^{\lambda_1}_\nu}{z_\nu} p_\nu
=\alpha^n  {a_{\lambda \mu}}^\nu \frac{ p_\nu}{z_\nu}  \, . \label{ProductAtOne}
\end{align}
Thus the product $\circ$ that we derived in equivariant cohomology indeed coincides with the convolution product inherited from class functions through the isomorphism  \cite{LS,FJW}
\begin{equation}
C_\mu   \leftrightarrow \frac{p_\mu}{z_\mu}
\, ,
\end{equation} 
up to an important extra constant factor  $\alpha^n$ in equation (\ref{ProductAtOne}).

\subsubsection{Normalization and Limit}
We still need to show that in the limit $\alpha=\beta \rightarrow 0$ the product is projected onto the {\em top} degree connection coefficients in order to find the ring structure of the ordinary cohomology.
To see how this comes about, we rescale the power sum symmetric functions to
$\tilde{p}_\lambda$:
\begin{equation}
p_\lambda = \alpha^{l(\lambda)} \tilde{p}_\lambda \, .
\end{equation}
As we will soon demonstrate, the rescaling is necessary to concentrate on the leading non-trivial structure constant in the $\alpha \rightarrow 0$ limit. 
We  find the relation:
\begin{equation}
\frac{\tilde{p}_\lambda}{z_\lambda} \circ \frac{\tilde{p}_\mu}{z_\mu} = \alpha^{n-l(\lambda)-l(\mu)+l(\nu)} {a_{\lambda \mu}}^\nu \frac{\tilde{p}_\nu}{z_\nu} \label{AlphaScaling} \, .
\end{equation}
Moreover, recall the degree $\text{deg}(\lambda)=n-l(\lambda)$ of a permutation $\lambda$. Thus, when the degrees of $\lambda,\mu$ add to the degree of the permutation $\nu$, we have the relation:
\begin{equation}
l(\lambda) + l(\mu) -l(\nu) = n \, .
\end{equation}
When we take the $\alpha \rightarrow 0$ limit, the dominant term in the right hand side of (\ref{AlphaScaling}) is the one with minimal power of $\alpha$ which is when the length of the result $l(\nu)$ is minimal, namely the degree of the resulting permutation $\nu$ is maximal. When the degree is maximal, we find that the $\alpha$ dependent factor drops out and  the multiplication indeed becomes the multiplication in the ordinary cohomology.  Only the top degree multiplication survives in the $\alpha \rightarrow 0$ limit. This shows that the equivariant cohomology ring  contains all information on the ordinary cohomology ring, as it must.

\subsubsection{The Farahat-Higman Ring and Its Polynomials}
Before we explore the information that the equivariant ring contains beyond the ordinary cohomology limit, we point out that taking the limit towards the ordinary cohomology theory provides a clear extra motivation for results  on the central algebra of functions on the group of permutations \cite{FarahatHigman} as well as the definition  of associated symmetric polynomials \cite{Macdonald} and the study of their properties \cite{GouldenJackson}. 
We recall that in a classic book in the field of symmetric polynomials \cite{Macdonald}, the ring of top coefficients is defined through the filtering by degree and by considering the standard construction of the ring that arises from the filtration\footnote{ Chapter I, paragraph 7, examples 24 and 25 of \cite{Macdonald}.}. This procedure precisely coincides with picking out the dominant term in the limiting procedure above.
Moreover, there turn out to be an associated set of polynomials, which we name Farahat-Higman-Macdonald polynomials and which capture the top connection coefficient structure constants. We review the definition and construction of these polynomials in Appendix \ref{gPolynomials} and provide a table of these polynomials for small symmetric groups $S_n$. These polynomials represent well the multiplication of the ordinary topologically twisted model on the Hilbert scheme of points   and have a mildly modified cousin that is stable under taking the large $n$ limit \cite{IvanovKerov,Li:2020zwo}.
This Farahat-Higman ring has an interesting generalization  to Hilbert schemes of orbifolds of $\mathbb{C}^2$. This ring and the ordinary cohomology of these spaces was analyzed in \cite{Wang}. 

In summary, we linked up the equivariant to the ordinary cohomology ring. Next, we return to the study of the equivariant structure constants.

\subsection{The Jack Characters and Structure Constants}
\label{JackCharactersStructureConstants}
We saw that the geometric  construction of a Heisenberg algebra \cite{NakajimaLectures,Grojnowski} gives rise to a power sum function basis of the cohomology while the fixed point basis relates naturally to the Jack functions $P_\lambda^{(\beta/\alpha)}$. It will be convenient in the following to rescale the latter functions to the integral form Jack functions $J_\lambda^{(\beta/\alpha)}$:
\begin{equation}
J_\lambda^{(\alpha)} = c_\lambda P_\lambda^{(\alpha)} \, ,
\end{equation}
where $c_\lambda(\alpha) = \prod_{s \in \lambda} (\alpha a(s) + l(s) +1) $ \cite{Macdonald}. We also define $c_\lambda'(\alpha)=\prod_{s \in \lambda} (\alpha a(s) + l(s)+\alpha)$.
The relation between the Jack and power sum  bases  gives rise to  basis change coefficients which are called Jack characters $\theta_\mu^\lambda(\alpha)$:
\begin{align}
J_\lambda^{(\alpha)} &= \theta_\mu^\lambda(\alpha) p_\mu \, .
\end{align}
Using the orthogonality relations \cite{Macdonald}:
\begin{align}
\sum_\rho z_\rho \alpha^{l(\rho)} \theta_\rho^\lambda(\alpha) \theta_\rho^\mu (\alpha) &= \delta_{\lambda \mu} c_\lambda(\alpha) c_\lambda'(\alpha) \nonumber \\
\sum_\lambda c_\lambda^{-1}(\alpha) (c'_\lambda(\alpha))^{-1} \theta_\rho^\lambda(\alpha) \theta_\sigma^\lambda(\alpha) &= \delta_{\rho \sigma} \frac{1}{z_\rho} \alpha^{-l(\rho)} 
\end{align}
we find the inverse relation:
\begin{align}
\frac{p_\rho}{z_\rho} & = \alpha^{l(\rho)} \frac{1}{c_\lambda'} \theta_\rho^\lambda P_\lambda^{(\alpha)} \, .
\end{align}
Thus, we can also compute the structure constants of the equivariant ring in terms of the Heisenberg algebra of generators of the cohomology:
\begin{align}
\frac{p_\lambda}{z_\lambda} \circ \frac{p_\mu}{z_\mu} &=(\frac{\beta}{\alpha})^{l(\lambda)}
\frac{\theta^{\lambda_1}_\lambda}{c_{\lambda_1}'} P_{\lambda_1}^{(\beta/\alpha)}
\circ (\frac{\beta}{\alpha})^{l(\mu)} \frac{\theta^{\lambda_2}_\mu}{c_{\lambda_2}'} P_{\lambda_2}^{(\beta/\alpha)}
\nonumber \\
&= \alpha^{n- l(\lambda) - l(\mu)}
\beta^{l(\lambda)+l(\mu)}
\theta^{\lambda_1}_\lambda \theta^{\lambda_1}_\mu  \frac{1}{c_{\lambda_1}'} P_{\lambda_1}^{(\beta/\alpha)}
\nonumber \\
&=  \alpha^{n} (\frac{\beta}{\alpha})^{l(\lambda)+l(\mu)}
\theta^{\lambda_1}_\lambda \theta^{\lambda_1}_\mu  \theta^{\lambda_1}_\nu \frac{1}{c_{\lambda_1} c_{\lambda_1}'} p_\nu \, .
\end{align}
The structure constants are closely related to a quantity called Jack connection coefficients. The latter are defined as follows \cite{GouldenJackson}. We recall the norm of the (integral form) Jack symmetric functions:
\begin{align}
\langle J_\lambda^{(\alpha)},J^{(\alpha)}_\mu \rangle_\alpha &= \delta_{\lambda \mu} c_\lambda(\alpha) c'_\mu(\alpha)
\, .
\end{align}
We then  have the definition of Jack connection coefficients ${c_{\pi \sigma}}^\lambda$ through the identification of the two generating series \cite{GouldenJackson}:
\begin{align}
\sum_{\theta} \frac{1}{\langle J_\theta^{(\alpha)}, J_\theta^{(\alpha)} \rangle_\alpha} J_\theta^{(\alpha)}(x) J_\theta^{(\alpha)}(y) J_\theta^{(\alpha)}(z) t^{|\theta|}
&= \sum_{\beta} \frac{1}{c_\beta c_\beta'} \theta^\beta_\pi p_\pi(x) \theta^\beta_\sigma p_\sigma(y) \theta^\beta_\lambda p_\lambda(z) t^{|\beta|} \nonumber \\
&=1+ \sum_{n \ge 1} t^n \sum_{\lambda,\pi,\sigma \vdash n} {c_{\pi,\sigma}}^\lambda \alpha^{-l(\lambda)}
z_\lambda^{-1} p_\pi(x) p_\sigma(y) p_\lambda(z) \, ,
\end{align}
from which we conclude that the Jack connection coefficients are:
\begin{align}
{c_{\pi,\sigma}}^\lambda &= \sum_\beta \alpha^{l(\lambda)} z_\lambda \frac{1}{c_\beta(\alpha) c_\beta'(\alpha)} \theta^\beta_\pi (\alpha) \theta^\beta_\sigma (\alpha) \theta^\beta_\lambda(\alpha) \, .
\label{JackConnectionCoefficients}
\end{align}
Thus we can write the   structure constants of the product $\circ$ as: 
\begin{align}
\frac{p_\lambda}{z_\lambda} \circ \frac{p_\mu}{z_\mu} 
&=  \alpha^{n} (\frac{\beta}{\alpha})^{l(\lambda)+l(\mu)-l(\nu)}
{c_{\lambda \mu}}^\nu \frac{p_\nu}{z_\nu} \, .
\end{align}
Firstly, to simplify the structure constants, we rescale $p_\lambda$ by the appropriate power of $\beta/\alpha$. Secondly, if we wish to expand around the structure constants of the ordinary cohomology, then we rescale the power sum $p_\lambda$ by the power of $\alpha$ determined previously. Thus, our rescaled description in terms of $\tilde{p}_\lambda = \beta^{l(\lambda)} p_\lambda$ reads:
\begin{align}
\frac{\tilde{p}_\lambda}{z_\lambda} \circ \frac{\tilde{p}_\mu}{z_\mu} 
&=  \alpha^{n-l(\lambda)-l(\mu)+l(\nu)} 
{c_{\lambda \mu}}^\nu  \frac{\tilde{p}_\nu}{z_\nu} \, .
\end{align}
This formula, together with the expression (\ref{JackConnectionCoefficients}) for the structure constants, is our most concrete  description of the product $\circ$ in equivariant cohomology. Some of the properties of the structure constants are manifest, while others are highly non-trivial. 

\subsubsection*{Properties of the Jack Structure Constants}
It is interesting to  study  the $\alpha \rightarrow 0$ limit once more. The leading term is the one with the top degree $n-l(\nu)$ for which the power of $\alpha$ up front annihilates. However, note that there is now a potential dependence of ${c_{\lambda \mu}}^\nu (\beta/\alpha)$ on the ratio $\alpha_J=\beta/\alpha=1+\gamma$ in the product $\circ$. Physically, we expect no such dependence in the $\alpha \rightarrow 0$ limit since the ordinary (non-equivariant) limit (at fixed $\beta/\alpha$) is unique. 

It is a non-trivial mathematical theorem that the top Jack connection coefficient is independent of $\beta/\alpha$. In fact, one can prove considerably stronger properties \cite{DolegaFeray}. 
The Jack structure constant ${c_{\pi,\sigma}}^\lambda$ is a polynomial in the variable $\gamma=\alpha_J-1$. Moreover, it is of maximal degree $d$ given by:
\begin{equation}
d = n-l(\pi)-l(\sigma)+l(\lambda) \, 
\end{equation}
in the deformation parameter $\gamma$. In particular, the top connection coefficient, though it naively can depend on $\gamma$, does not because it is a polynomial in $\gamma$ of order zero. This is in accord with the physical intuition that the non-equivariant limit is unique. 
The theorem \cite{DolegaFeray} is  more powerful. 

A second observation is that the structure constants satisfy a duality property. We expect a symmetry between $\alpha$ and $\beta$ (from interchanging the directions $(x,y)$ in the original complex two-plane). Indeed, the structure constants satisfy the duality formula  \cite{Vassilieva}:
\begin{align}
{c_{\pi,\sigma}}^\lambda (\alpha_J^{-1}) &= \sum_\beta \alpha_J^{-l(\lambda)} z_\lambda \frac{1}{c_\beta(\alpha_J^{-1}) c_\beta'(\alpha_J^{-1})} \theta^\beta_\pi  (\alpha_J^{-1}) \theta^\beta_\sigma  (\alpha_J^{-1})  \theta^\beta_\lambda (\alpha_J^{-1}) 
\nonumber \\
&= (-\alpha_J)^{l(\pi)+l(\sigma)+l(\lambda)-3n} 
 \sum_\beta \alpha_J^{-l(\lambda)} z_\lambda \frac{\alpha_J^{2 n}}{c_\beta(\alpha_J) c_\beta'(\alpha_J)} \theta^{\beta}_\pi  (\alpha_J) \theta^{\beta}_\sigma  (\alpha_J)  \theta^{\beta}_\lambda (\alpha_J)
 \nonumber \\
&= (-\alpha_J)^{l(\pi)+l(\sigma)-l(\lambda)-n} 
 {c_{\pi,\sigma}}^\lambda(\alpha_J)
 \, .
\end{align}
This formula captures the $\alpha \leftrightarrow \beta$ duality of the equivariant cohomology ring which follows from the symmetry of the initial two planes $\mathbb{C}$ in the original space $\mathbb{C}^2$. 

Thirdly, it was conjectured that the Jack structure constants are polynomials in $\gamma$ with {\em positive integer} coefficients \cite{GouldenJackson}. Meanwhile, it has been proven that they are indeed polynomial \cite{DolegaFeray} with integer coefficients \cite{BenDali} and that the leading coefficient of the polynomial is a positive integer  \cite{Burchardt,BurchardtPhD}. The fully general conjecture (including the search for a combinatorial interpretation in terms of matchings \cite{GouldenJackson}) remains open. It seems to indicate the possibility of a much neater interpretation of the coefficients as counting the dimension of an appropriate space (namely, a categorification).

 While there is  no closed formula for the Jack structure constants, there are a few concrete general results \cite{GouldenJackson}. For instance, the coefficient of the partition $[1^n]$ in a generic product of partitions is known: 
 \begin{equation}
 {c_{\mu \nu}}^{[1^n]} = \delta_{\mu \nu} |C_\mu| (1+\gamma)^{n-l(\mu)} \, .
 \end{equation}
 Moreover, there is a sum rule:
 \begin{align}
 \sum_{\nu}  {c_{\mu \nu}}^{\lambda} = |C_\mu| (1+\gamma)^{n-l(\mu)}
  \, .
 \end{align}
 There is also a known generating series for the coefficient of the partition $[2 1^n]$ in a generic product. See e.g. \cite{GouldenJackson} and \cite{Vassilieva} for further results.
 
 In Appendix \ref{JackStructureConstants}
 we  tabulate all $S_n$ Jack structure constants for $n=2,3,4,5$ and give a few more results at higher $n$. For $n=2,3,4$, these formulas coincide with those of \cite{GouldenJackson}. With present computer power, these results can be extended considerably if desired. On these particular examples, it is easy to check that the above theorems and conjectures hold. 
Moreover, the tables contain calculations beyond those implied by the general results. It should be clear that these structure constants characterize the operator ring of the cohomology of the physical models, in the case of the equivariant Hilbert scheme target $\text{Hilb}^n(\mathbb{C}^2)$.

 It will be interesting to discuss the large $n$ limit of these structure constants in more detail. The large $n$ limit is quite natural from a  gauge theory perspective (in which the $S_n$ gauge symmetry is thought of as a remnant of a $U(n)$ gauge symmetry) and we expect the structure constants to stabilize at large $n$ when expressed in the appropriate way. We know that this is the case in the non-equivariant setting \cite{IvanovKerov,Li:2020zwo}. A relevant result may be the approximate factorization of appropriate cumulants of generalized characters at large $n$ \cite{Sniady}. We  very briefly comment on this issue in Appendix \ref{Largen}. 
Finally, there are further interesting results on polynomiality and other characteristics when one rewrites the structure constants in terms of the (Liouville) variable 
$\delta = \sqrt{\alpha_J} - \sqrt{\alpha_J}^{-1}$ \cite{Sniady}. Both the large $n$ behavior and these results  deserve to be  understood better.

\subsection*{Remarks}
To conclude this section, we offer a few questions and an observation.
\begin{itemize}

\item
An intriguing question is whether an index can be defined that would help in deriving the positive integrality of the structure constants. 
\item It would be interesting to  characterize the large $n$ limit of the Jack structure constants.
\item Recall that in section \ref{EquivariantLandscape} we used the chiral boson operator  $\alpha_{-n}$ to generate the state we coded in the power sum symmetric function $p_n$.
When the equivariant parameters satisfy $t_2/t_1=-1$, we have that the transposition class corresponds to   a cubic interaction of the type $(\partial X)^3$ \cite{FW} which is an interesting chiral perturbation of a free scalar field $X$ \cite{Dijkgraaf:1996iy}, related to two-dimensional Yang-Mills theory \cite{Douglas:1993wy}. It is the set-up where splitting and joining are equally strong. 
These observations make it possible to think of the classical twist interaction $[2,1^{n-2}]$ as a cubic deformation of a chiral boson.
It may then be natural to ask which states (or symmetric functions) diagonalize the  cubic interacting Hamiltonian. The answer to this is a standard part of symmetric function theory. The classical equivariant cubic operator  is diagonalized by Jack symmetric polynomials with known eigenvalues. Indeed, the cubic Hamiltonian is equivalent to a known integrable system, a Calogero-Moser Hamiltonian. 
Given these observations, it may be possible to understand the equivariant cohomology  as an integrable deformation of a free model.

\end{itemize}

\section{Conclusions}
\label{Conclusions}

In this paper, we proposed to embed a standard $AdS_3/CFT_2$ correspondence into a space of equivariant deformations. At string radius, the operator algebra of the one half BPS states contains a subalgebra (obtained by modding out by the non-trivial topology of $M$ in $AdS_3 \times S^3 \times M$) that coincides with the cohomology of the Hilbert scheme of the complex plane $\mathbb{C}^2$ (or the orbifold cohomology of the symmetric orbifold thereof). We viewed equivariant Gromov-Witten theories on $\mathbb{P}^1$ or $\mathbb{P}^1 \times \mathbb{C}^2$ as providing more general gauge/gravity correspondences. We analyzed the equivariant deformation of the Gromov-Witten/Hurwitz correspondence and showed that it allows to interpolate between a product of two pure topological gravity theories and the non-equivariant theory with matter. This provides a  model for how to couple topological quantum gravities consistently using world sheet instantons. We also explored the equivariant deformations of the standard cohomology of the Hilbert scheme of points on $\mathbb{C}^2$. We connected the geometric and algebraic literature such that the link between the equivariant cohomology and the ordinary cohomology becomes calculationally manifest. We reviewed intriguing integrality and positivity properties of the full equivariant structure constants. All these analyses can be viewed as  initial explorations 
of the equivariant space of couplings near points where we understand the gauge/gravity duality rather well. We believe the proposed framework allows for further fruitful steps in our understanding of holography.

\section*{Acknowledgments}
It is a pleasure to thank all my colleagues for creating a stimulating work environment.  

\appendix

\section{Permutations and Symmetric Functions}
\label{SymmetricPolynomials}
\label{Symmetric}
In this Appendix, we first recall elementary  facts in the theory of symmetric functions. For a full description we refer to \cite{Macdonald}. We gather a bestiary that should be useful to most readers.

\subsection{Permutations and Partitions}
The symmetric group $S_n$ has group elements $\pi$ that we can think of as permuting $n$ objects. The conjugacy class of a permutation in the symmetric group $S_n$ is characterized by a partition of $n$ that corresponds to the length of the orbits of the permutation. A transposition is the group element that exchanges two elements only. The length $l(\pi)$ of a partition $\pi$ equals the number of its non-zero parts. 

%
 
 We will describe partitions $\lambda$ by  positive integer numbers $\lambda_1 \ge \lambda_2 \ge \dots$ with finitely many non-zero entries. The parts of the partition are the non-zero entries. 
 The sum of the parts is the weight $|\lambda|=n$. Another notation for a partition $\lambda$ is $\lambda=[1^{m_1} 2^{m_2} \dots]$.\footnote{When the meaning is clear, we sometimes omit the square brackets.} The multiplicity of  the number $i$ in the partition $\lambda$ is $m_i=m_i(\lambda)$.  
 We can associate a Young diagram $D_\lambda \equiv \lambda$ to the partition $\lambda$ with $\lambda_1$ boxes in the top line, $\lambda_2$ boxes in the next (left-aligned) line, et cetera. The conjugate partition $\lambda'$ corresponds to the transposed diagram. 
 The hook length of $\lambda$ at the position $x=(i,j)$ in the diagram of $\lambda$ equals $h(i,j)=\lambda_i+\lambda_j'-i-j+1$. 
 We can define the arm and leg of a box $(i,j)$.  We have the arm $a(i,j)=\lambda_i-j$ and leg $l(i,j)=\lambda_j'-i$. 
Therefore, we have the hook length $h(i,j)=a(i,j)+l(i,j)+1$.
We  define the product of hook lengths of a diagram: $h_\lambda = \prod_{s \in \lambda}h(s)$. 
We also define the parameter dependent generalizations $c_\lambda(\alpha)$ and $c'_\lambda(\alpha)$ \cite{Macdonald}:
\begin{align}
c_\lambda(\alpha) &= \prod_{s \in \lambda} (\alpha a(s) + l(s) + 1)
\nonumber \\
c_{\lambda}'(\alpha) &= \prod_{s \in \lambda} (\alpha a(s) + l(s) + \alpha) \, .
\end{align}
In the bulk of the paper, we also use the alternative notations \cite{LQW}:
\begin{align}
\tilde{c}_\lambda(\alpha,\beta) &= \alpha^n {c}_{\lambda}(\beta/\alpha)
\nonumber \\
\tilde{c}_{\lambda}'(\alpha,\beta) &= \alpha^n c_{\lambda}'(\beta/\alpha) \, .
\end{align}
We define the useful number:
\begin{equation}
z(\lambda)=z_\lambda = \prod_{i \ge 1} i^{m_i} m_i! \, .
\end{equation}
The number $|C_\lambda|$  of symmetric group elements in the conjugacy class $\lambda$ equals $n!/z_\lambda$. 

\subsection{Symmetric Functions}
We  also need statements in symmetric function theory. In particular, we  use the definition of the symmetric monomials and the Schur, complete homogeneous and power sum symmetric functions.
Their relations and definitions are as follows.
The symmetric polynomials $\Lambda_n=\mathbb{Z}[x_1,\dots,x_n]^{S_n}$ are a graded subring of polynomials. We use the notation $x^\alpha=x_1^{\alpha_1} \dots x_n^{\alpha_n}$.  Let $\lambda$ be a partition. We then define the symmetric monomials
\begin{equation}
m_\lambda(x_i) = \sum_{\alpha \, \text{perm. of} \,  \lambda} x^{\alpha} \, .
\end{equation}
As $\lambda$ runs through partitions of length $\le n$, the $m_\lambda$ form a basis of the symmetric polynomials $\Lambda_n$. 
The number of variables is often irrelevant, as long as it is large enough. We call them symmetric functions when there are an infinite number of variables \cite{Macdonald}. 
The {complete homogeneous} symmetric function $h_r$ is:
\begin{equation}
h_r = \sum_{|\lambda|=r} m_\lambda
\, .
\end{equation}
A generating function for the complete homogeneous symmetric functions is:
\begin{equation}
H(t) = \sum_{r \ge 0} h_r t^r  = \prod_{i \ge 1} (1-x_i t)^{-1} \, .
\end{equation}
For a more general partition $\lambda$, we put $h_\lambda=h_{\lambda_1} h_{\lambda_2} \dots$ 
The power sum symmetric functions are defined as:
\begin{equation}
p_r = \sum x_i^r = m_{r} \, ,
\end{equation}
and for a more general partition we  again have the product formula $p_\lambda=p_{\lambda_1} p_{\lambda_2} \dots$
A possible definition of the Schur functions is in terms of the complete homogeneous symmetric functions:
\begin{equation}
s_\lambda = \det (h_{\lambda_i-i+j})_{1 \le i,j \le n} \, 
\end{equation} with $n$ large enough. 
The transition matrix 
to change from power sum to Schur symmetric functions is the character table of the symmetric group $S_n$. 

The Jack functions $P^{(\alpha)}_\lambda$ are characterized by the fact that the transition matrix from Jack functions to the monomials  is strictly upper unitriangular and that they are  pairwise orthogonal with respect to the scalar product 
\begin{equation}
\langle p_\lambda , p_\mu \rangle = \alpha^{l(\lambda)} z_\lambda \delta_{\lambda \mu} \, .
\end{equation}
They capture classes of symmetric functions through the specialization of the parameter $\alpha$: $P^{(1)}_\lambda=s_\lambda, P^{(0)}_\lambda=e_\lambda$ and $P^{(\infty)}_\lambda=m_\lambda$. 
There is a second standard normalization which reads:
\begin{equation}
J_\lambda^{(\alpha)} = c_\lambda(\alpha) P_\lambda^{(\alpha)} 
\, .
\end{equation}
For concreteness and convenience, we provide a table of these Jack symmetric  functions $J$ for small partitions in Table \ref{JackSymmetricFunctions}.
\begin{table}
\begin{center}
\begin{tabular}{ |c|c| } 
 \hline 
 $\lambda$ & $J_\lambda^{(\alpha)} $\\
 \hline 
 $[1]$ & $m_1$ \\ 
 $[2]$ & $(1+\alpha)m_2+2m_{11}$  \\ 
 $[1^2]$ & $2 m_{11}$  \\
 ${[3]} $ & $(\alpha +1) \left((2 \alpha +1) m_3+3 m_{21}\right)+6 m_{111} $ \\
 $[2,1]$ & $ (\alpha +2) m_{21}+6 m_{111}$ \\
 $[1^3]$ & $ 6 m_{111}$ \\
 $[{4}]$ & 
 \makecell{
 $(\alpha +1) (2 \alpha +1) (3 \alpha +1) m_4+2 (\alpha +1) (3 (\alpha +1) m_{22}$ \\$ +(4 \alpha +2) m_{31}+6 m_{211})+24 m_{1111}$} \\
 ${[3,1]}$ & $2 \left((\alpha +1)^2 m_{31}+2 (\alpha +1) m_{22}+(3 \alpha +5) m_{211}+12 m_{1111}\right)$ \\
$ [{2,2}]$ & $2 (\alpha +1) (\alpha +2) m_{22}+4 (\alpha +2) m_{211}+24 m_{1111}$ \\
 $[{2,1^2}]$&  $2 (\alpha +3) m_{211}+24 m_{1111}$ \\
 $[1^4]$ & $24 m_{1111}$ \\
 \hline
\end{tabular}
\end{center}
\caption{The Jack symmetric functions $J_\lambda^{(\alpha)}$ in the monomial basis for partitions $\lambda$ of $1,2,3$ and $4$.}
\label{JackSymmetricFunctions}
\end{table}

\subsection{The Farahat-Higman-Macdonald Polynomials}
\label{gPolynomials}
This subsection is dedicated to making a classic construction\footnote{Examples 24 and 25, paragraph 7, chapter 1 of \cite{Macdonald}.} of polynomials relevant to the structure constants of the ordinary cohomology of the Hilbert scheme of points on $\mathbb{C}^2$ sufficiently concrete to provide a table of such polynomials. 

Firstly, we need to define a modified cycle type. It takes a partition $\lambda$ representing a cycle and subtracts $1$ from each entry. The modified cycle type has the advantage of being stable  under the embedding $S_n \rightarrow S_{n+1}$. Secondly, we denote $c_\lambda(n)$ the sum of those $w \in S_n$ with modified cycle type $\lambda$.\footnote{We  use the notation of \cite{Macdonald} here for ease of reference.} We moreover introduce the structure constants $c_\lambda c_\mu = {{{b}_{\lambda \mu}}}^\nu (n) c_\nu$. This ring is filtered by a degree $|\lambda|$. We can divide out terms in the product that have lower degree and in this manner obtain the  associated graded ring in which:
\begin{equation}
c_\lambda c_\mu = \sum_{|\nu|=|\lambda|+|\mu|} {{{b}_{\lambda \mu}}}^\nu (n) c_\nu
\, .
\end{equation}
This is the Farahat-Higman ring \cite{FarahatHigman}.
We recall that this equation has labels corresponding to the modified cycle type. 

Following \cite{Macdonald}, we can associate polynomials $g_\lambda$ to this ring.  Firstly, we define the polynomials $h^\ast$ dual under an appropriate involution to the complete homogeneous polynomials $h$. We invert the generating function:
\begin{equation}
u = t H(t) = t+ h_1 t^2 + \dots
\end{equation}
for the variable $t$ as a function of $u$:
\begin{equation}
t = u + h_1^\ast u^2 + h_2^\ast u^3 + \dots
\end{equation}
to define polynomials $h_n^\ast$ and furthermore define the products
\begin{equation}
h_\lambda^\ast = h_{\lambda_1}^\ast h_{\lambda_2}^\ast \dots
\, .
\end{equation}
The polynomials $h_\lambda^\ast$ form a basis of the symmetric functions. The  polynomials $g_\lambda$ dual to the polynomials $h_\lambda^\ast$ with respect to the standard inner product are what we call the Farahat-Higman-Macdonald polynomials $g_\lambda$. One can show that these are triangular combinations of the monomials $m_\mu$:
\begin{equation}
g_\lambda = (-1)^{l(\lambda)} m_\lambda + \sum_{\mu > \lambda} u_{\mu \lambda} m_\mu
\, .
\end{equation}
It is proven in \cite{Macdonald,GouldenJacksonMacdonald} that these polynomials $g_\lambda$ multiply as do the conjugacy classes $c_\lambda$ at the top degree (where both are labelled by modified partitions). To add to the concrete description of the ordinary cohomology ring discussed at greater length in 
\cite{Li:2020zwo}, we can thus table the symmetric polynomials $g_\lambda$ in Table \ref{FarahatHigmanMacdonaldPolynomials} and note that their multiplication captures the top connection coefficients that played the central role in \cite{LS,Li:2020zwo}. 
We generated  Table \ref{FarahatHigmanMacdonaldPolynomials}
 using Mathematica, the package Symmetric Functions \cite{Alexandersson} and very limited further programming.
 A table with the polynomials for partitions of up to $n=4$ is provided in \cite{GouldenJacksonMacdonald}.
\begin{table}
\begin{center}
\begin{tabular}{ |c|c| } 
 \hline 
 Partition $\lambda$ & $g_\lambda  $\\
 \hline 
 $[1]$ & $-p_1$ \\ 
 $[2]$ & $-p_2$  \\ 
 $[1^2]$ & $\frac{3 p_ 2}{2}+\frac{p_{11}}{2}$  \\
 ${[3]} $ & $-p_3 $ \\
 ${[2,1]}$ & $ 4 p_ 3+p_{21}$ \\
 ${[1^3]}$ & $ -\frac{10 p_ 3}{3}-\frac{3 p_{21}}{2}-\frac{p_{111}}{6}$ \\
 ${[4]}$ &  $-p_4$ \\
 ${[3,1]}$ & $5 p_ 4+p_{31}$ \\
 $[{2^2}]$ & $\frac{5 p_ 4}{2}+\frac{p_{22}}{2}$ \\
 $[{2,1^2}]$&  $-15 p_ 4-4 p_{31}-\frac{3 p_{22}}{2}-\frac{p_{211}}{2}$ \\
 $[1^4]$ & $\frac{35 p_ 4}{4}+\frac{10 p_{31}}{3}+\frac{9 p_{22}}{8}+\frac{3 \
p_{211}}{4}+\frac{p_{1111}}{24}$ \\
 $[{5}]$ &  $-p_5$ \\
 $[4,1]$ & $6 p_ 5+p_{41}$ \\
 $[3,2]$ & 
$ 6 p_ 5+p_{32}$ \\
$[3,1^2]$ & $-21 p_ 5-\frac{3 p_{32}}{2}-5 
p_{41}-\frac{p_{311}}{2} $\\
$[2^2,1]$ & $-21 p_ 5-4 p_{32}-\frac{5 
p_{41}}{2}-\frac{p_{221}}{2} $\\
$[2,1^3]$ & $56 p_ 5+\frac{28 p_{32}}{3}+15 
p_{41}+\frac{3 p_{221}}{2}+2 p_{311}+\frac{p_{2111}}{6}
$\\
 $[1^5]$ & $-\frac{126 p_
5}{5}-5 p_{32}-\frac{35 p_{41}}{4}-\frac{9 p_{221}}{8}-\frac{5 \
p_{311}}{3}-\frac{p_{2111}}{4}-\frac{1}{120} p_{11111}$ \\
 \hline
\end{tabular}
\end{center}
\caption{Farahat-Higman-Macdonald symmetric polynomials  in the power sum basis up to partitions of five.}
\label{FarahatHigmanMacdonaldPolynomials}
\end{table}

\subsection{The Jack Structure Constants}
\label{JackStructureConstants}
In this subsection of the Appendix, we gather Jack connection coefficients for small orders of $S_n$ and a few useful illustrative results at higher $n$. One can easily generate more results than fit on a page by computer. The results were checked for $n \le 4$ against tables in \cite{GouldenJackson}. 

Firstly, for $n=1$, the only Jack connection coefficient is ${c_{1,1}}^1=1$. For $n=2$, we find table \ref{JackStructureConstantsAtn=2}.
\begin{table}
\begin{center}
\begin{tabular}{ |c|c|c| } 
 \hline 
 Partitions $\lambda_1 ; \lambda_2$ and $\lambda_3$ & $2$ & $1,1$ \\ 
 \hline
$2;2$ & $\gamma$ & $1+ \gamma$ \\
$2;1,1$ & $1$ & $0$ \\
$1,1;1,1$ & $0$ & $1$ \\
 \hline
\end{tabular}
\end{center}
\caption{The Jack connection coefficients and equivariant structure constants ${c_{\lambda_1,\lambda_2}}^{\lambda_3}$ for partitions of $2$.}
\label{JackStructureConstantsAtn=2}
\end{table}
For the group $S_3$, we  provide the  table \ref{JackStructureConstantsAtn=3}.
\begin{table}
\begin{center}
\begin{tabular}{ |c|c|c|c| } 
 \hline 
 Partitions $\lambda_1 ; \lambda_2$ and $\lambda_3$ & $3$ & $2,1$ & $1,1,1$ \\
 \hline
$3;3$ & $1+\gamma+2 \gamma^2$ & $2 \gamma+2 \gamma^2$ & $2+4 \gamma+2 \gamma^2$\\
$3;2,1$ & $3 \gamma$ & $2+2 \gamma$ & $0$ \\
$3;1,1,1$ & $1$ & $0$  & $0$ \\
$2,1;2,1$ & $3$ & $\gamma$ & $3 + 3 \gamma$ \\
$2,1;1,1,1$ & $0$ & $1$ & $0$ \\
$1,1,1;1,1,1$ & $0$ & $0$ & $1$ \\
 \hline
\end{tabular}
\end{center}
\caption{The Jack connection coefficients and equivariant structure constants for partitions of $3$.}
\label{JackStructureConstantsAtn=3}
\end{table}
For the value $n=4$, we   find the table \ref{JackStructureConstantsAtn=4} and for the group $S_5$ we have the tables \ref{JackStructureConstantsAtn=5Part1} and \ref{JackStructureConstantsAtn=5Part2}. These tables illustrate that we can be very concrete about the structure constants of the equivariant cohomology ring. Note in particular the positive integer coefficients in all  polynomials in $\gamma$. 
\begin{table}
\begin{center}
\begin{tabular}{ |c|c|c|c|c|c| }
 \hline 
  $\lambda_1 ; \lambda_2$ and $\lambda_3$ & $4$ & $3,1$ & $2,2$ & $2,1,1$ & $1,1,1,1$ \\
 \hline
 4 ; 4 & 
 \makecell{ $ 6 \gamma ^3+7 \gamma ^2 $ \\ $+7 \gamma $ }
 & 
 \makecell{ $ 6 \gamma ^3+9 \gamma ^2 $ \\$+6 \gamma +3 $ }
 & 
 \makecell{ $ 6 \gamma ^3+8 \gamma ^2 $ \\ $+4 \gamma +2 $ }
 & 
\makecell{ $ 6 \gamma ^3+12 \gamma ^2$ \\$+6 \gamma $} 
 & 
 \makecell{
 $ 6 \gamma ^3+18 \gamma ^2 $ \\ $+18 \gamma +6 $ }
 \\
 
 4 ; 3,1 & 
 $ 8 \gamma ^2+4 \gamma +4 $ 
 & 
 $ 6 \gamma ^2+6 \gamma $ 
 & 
 $ 8 \gamma ^2+8 \gamma $ 
 & 
 $ 4 \gamma ^2+8 \gamma +4 $ 
 & 
 $ 0 $ 
\\
 4 ; 2,2 & 
 $ 3 \gamma ^2+\gamma +1 $ 
 & 
 $ 3 \gamma ^2+3 \gamma $ 
 & 
 $ 2 \gamma ^2+2 \gamma $ 
 & 
 $ 2 \gamma ^2+4 \gamma +2 $ 
 & 
 $ 0 $ 
\\
 4 ; 2,1,1 & 
 $ 6 \gamma $ 
 & 
 $ 3 \gamma +3 $ 
 & 
 $ 4 \gamma +4 $ 
 & 
 $ 0 $ 
 & 
 $ 0 $ 
\\
 4 ; 1,1,1,1 & 
 $ 1 $ 
 & 
 $ 0 $ 
 & 
 $ 0 $ 
 & 
 $ 0 $ 
 & 
 $ 0 $ 
\\
 3,1 ; 3,1 & 
 $ 8 \gamma $ 
 & 
 $ 2 \gamma ^2+4 \gamma +4 $ 
 & 
 $ 8 \gamma +8 $ 
 & 
 $ 4 \gamma ^2+4 \gamma $ 
 & 
 $ 8 \gamma ^2+16 \gamma +8 $ 
\\
 3,1 ; 2,2 & 
 $ 4 \gamma $ 
 & 
 $ 3 \gamma +3 $ 
 & 
 $ 0 $ 
 & 
 $ 0 $ 
 & 
 $ 0 $ 
\\
 3,1 ; 2,1,1 & 
 $ 4 $ 
 & 
 $ 3 \gamma $ 
 & 
 $ 0 $ 
 & 
 $ 4 \gamma +4 $ 
 & 
 $ 0 $ 
\\
 3,1 ; 1,1,1,1 & 
 $ 0 $ 
 & 
 $ 1 $ 
 & 
 $ 0 $ 
 & 
 $ 0 $ 
 & 
 $ 0 $ 
\\
 2,2 ; 2,2 & 
 $ \gamma $ 
 & 
 $ 0 $ 
 & 
 $ \gamma ^2+2 \gamma +2 $ 
 & 
 $ \gamma ^2+\gamma $ 
 & 
 $ 3 \gamma ^2+6 \gamma +3 $ 
\\
 2,2 ; 2,1,1 & 
 $ 2 $ 
 & 
 $ 0 $ 
 & 
 $ 2 \gamma $ 
 & 
 $ \gamma +1 $ 
 & 
 $ 0 $ 
\\
 2,2 ; 1,1,1,1 & 
 $ 0 $ 
 & 
 $ 0 $ 
 & 
 $ 1 $ 
 & 
 $ 0 $ 
 & 
 $ 0 $ 
\\
 2,1,1 ; 2,1,1 & 
 $ 0 $ 
 & 
 $ 3 $ 
 & 
 $ 2 $ 
 & 
 $ \gamma $ 
 & 
 $ 6 \gamma +6 $ 
\\
 2,1,1 ; $1^4$ & 
 $ 0 $ 
 & 
 $ 0 $ 
 & 
 $ 0 $ 
 & 
 $ 1 $ 
 & 
 $ 0 $ 
 \\
$1^4$ ; $1^4$ & 
 $ 0 $ 
 & 
 $ 0 $ 
 & 
 $ 0 $ 
 & 
 $ 0 $ 
 & 
 $ 1 $ 
\\
 \hline
\end{tabular}
\end{center}
\caption{The Jack connection coefficients and equivariant structure constants for partitions of $4$.}
\label{JackStructureConstantsAtn=4}
\end{table}

\begin{table}
\begin{center}
\begin{tabular}{ |c|c|c|c| }
 \hline 
  $\lambda_1 ; \lambda_2$ and $\lambda_3$ & $5$ & $4,1$ & $3,2$   \\
 \hline
 5 ; 5 & 
\makecell{ $ 24 \gamma ^4+46 \gamma ^3$ \\ $+54 \gamma ^2+16 \gamma +8 $ }
 & 
\makecell{ $ 24 \gamma ^4+52 \gamma ^3$ \\$+56 \gamma ^2+28 \gamma $} 
 & 
 $ 24 \gamma ^4+48 \gamma ^3+48 \gamma ^2+24 \gamma $ 
\\
 5 ; 4,1 & 
 $ 30 \gamma ^3+35 \gamma ^2+35 \gamma $ 
 & 
 $ 24 \gamma ^3+36 \gamma ^2+24 \gamma +12 $ 
 & 
 $ 30 \gamma ^3+42 \gamma ^2+24 \gamma +12 $ 
\\
 5 ; 3,2 & 
 $ 20 \gamma ^3+20 \gamma ^2+20 \gamma $ 
 & 
 $ 20 \gamma ^3+28 \gamma ^2+16 \gamma +8 $ 
 & 
 $ 18 \gamma ^3+24 \gamma ^2+12 \gamma +6 $ 
\\
 5 ; 3,1,1 & 
 $ 20 \gamma ^2+10 \gamma +10 $ 
 & 
 $ 12 \gamma ^2+12 \gamma $ 
 & 
 $ 18 \gamma ^2+18 \gamma $ 
\\
 5 ; 2,2,1 & 
 $ 15 \gamma ^2+5 \gamma +5 $ 
 & 
 $ 12 \gamma ^2+12 \gamma $ 
 & 
 $ 12 \gamma ^2+12 \gamma $ 
\\
 5 ; 2,1,1,1 & 
 $ 10 \gamma $ 
 & 
 $ 4 \gamma +4 $ 
 & 
 $ 6 \gamma +6 $ 
\\
 5 ; 1,1,1,1,1 & 
 $ 1 $ 
 & 
 $ 0 $ 
 & 
 $ 0 $ 
%
\\
 4,1 ; 4,1 & 
 $ 30 \gamma ^2+15 \gamma +15 $ 
 & 
 $ 6 \gamma ^3+23 \gamma ^2+23 \gamma $ 
 & 
 $ 30 \gamma ^2+30 \gamma $ 
\\
 4,1 ; 3,2 & 
 $ 25 \gamma ^2+10 \gamma +10 $ 
 & 
 $ 20 \gamma ^2+20 \gamma $ 
 & 
 $ 18 \gamma ^2+18 \gamma $ 
\\
 4,1 ; 3,1,1 & 
 $ 15 \gamma $ 
 & 
 $ 8 \gamma ^2+8 \gamma +8 $ 
 & 
 $ 12 \gamma +12 $ 
\\
 4,1 ; 2,2,1 & 
 $ 15 \gamma $ 
 & 
 $ 3 \gamma ^2+9 \gamma +9 $ 
 & 
 $ 6 \gamma +6 $ 
\\
 4,1 ; 2,1,1,1 & 
 $ 5 $ 
 & 
 $ 6 \gamma $ 
 & 
 $ 0 $ 
\\
 4,1 ; $1^5$ & 
 $ 0 $ 
 & 
 $ 1 $ 
 & 
 $ 0 $ 
\\
 3,2 ; 3,2 & 
 $ 15 \gamma ^2+5 \gamma +5 $ 
 & 
 $ 12 \gamma ^2+12 \gamma $ 
 & 
 $ 2 \gamma ^3+13 \gamma ^2+13 \gamma $ 
\\
 3,2 ; 3,1,1 & 
 $ 15 \gamma $ 
 & 
 $ 8 \gamma +8 $ 
 & 
 $ 2 \gamma ^2+7 \gamma +7 $ 
\\
 3,2 ; 2,2,1 & 
 $ 10 \gamma $ 
 & 
 $ 4 \gamma +4 $ 
 & 
 $ 3 \gamma ^2+6 \gamma +6 $ 
\\
 3,2 ; 2,1,1,1 & 
 $ 5 $ 
 & 
 $ 0 $ 
 & 
 $ 4 \gamma $ 
\\
 3,2 ; $1^5$ & 
 $ 0 $ 
 & 
 $ 0 $ 
 & 
 $ 1 $ 
\\
 3,1,1 ; 3,1,1 & 
 $ 5 $ 
 & 
 $ 8 \gamma $ 
 & 
 $ 0 $ 
\\
 3,1,1 ; 2,2,1 & 
 $ 5 $ 
 & 
 $ 4 \gamma $ 
 & 
 $ 3 \gamma $ 
\\
 3,1,1 ; $2,1,1,1$ & 
 $ 0 $ 
 & 
 $ 4 $ 
 & 
 $ 1 $ 
\\
 3,1,1 ; $1^5$ & 
 $ 0 $ 
 & 
 $ 0 $ 
 & 
 $ 0 $ 
\\
 2,2,1 ; 2,2,1 & 
 $ 5 $ 
 & 
 $ \gamma $ 
 & 
 $ 3 \gamma $ 
\\
 2,2,1 ; 2,1,1,1 & 
 $ 0 $ 
 & 
 $ 2 $ 
 & 
 $ 3 $ 
\\
 2,2,1 ; $1^5$ & 
 $ 0 $ 
 & 
 $ 0 $ 
 & 
 $ 0 $ 
\\
 2,1,1,1 ; 2,1,1,1 & 
 $ 0 $ 
 & 
 $ 0 $ 
 & 
 $ 0 $ 
\\
 2,1,1,1 ; $1^5$ & 
 $ 0 $ 
 & 
 $ 0 $ 
 & 
 $ 0 $ 
\\
 $1^5$ ; $1^5$ & 
 $ 0 $ 
 & 
 $ 0 $ 
 & 
 $ 0 $ 
\\
 \hline
\end{tabular}
\end{center}
\caption{The Jack connection coefficients and equivariant structure constants for partitions of $5$, part 1.}
\label{JackStructureConstantsAtn=5Part1}
\end{table}

\begin{table}
\begin{center}
\begin{tabular}{ |c|c|c|c|c|c|}
 \hline 
  $\lambda_1 ; \lambda_2$ and $\lambda_3$ & $3,1,1$ & $2,2,1$ & $2,1^3$ & $1^5$   \\
 \hline
  5 ; 5 & 
  \makecell{
 $ 24 \gamma ^4+60 \gamma ^3
 $ \\ $+60 \gamma ^2+36 \gamma +12 $ }
 & \makecell{
 $ 24 \gamma ^4+56 \gamma ^3$ \\ $+48 \gamma ^2+24 \gamma +8 $ }
 &  \makecell{
 $ 24 \gamma ^4+72 \gamma ^3$ \\ $+72 \gamma ^2+24 \gamma $ }
 & \makecell{
 $ 24 \gamma ^4+96 \gamma ^3$ \\ $+144 \gamma ^2+96 \gamma $ \\$ +24 $} 
\\
 5 ; 4,1 & 
 \makecell{
 $ 18 \gamma ^3+36 \gamma ^2$ \\ $+18 \gamma $ }
 & 
 $ 24 \gamma ^3+48 \gamma ^2+24 \gamma $ 
 & \makecell{
 $ 12 \gamma ^3+36 \gamma ^2 $ \\ $+36 \gamma +12 $ }
 & 
 $ 0 $ 
\\
 5 ; 3,2 & 
 $ 18 \gamma ^3+36 \gamma ^2+18 \gamma $ 
 & 
 $ 16 \gamma ^3+32 \gamma ^2+16 \gamma $ 
 & \makecell{
 $ 12 \gamma ^3+36 \gamma ^2$ \\ $+36 \gamma +12 $ }
 & 
 $ 0 $ 
\\
 5 ; 3,1,1 & 
 $ 6 \gamma ^2+12 \gamma +6 $ 
 & 
 $ 8 \gamma ^2+16 \gamma +8 $ 
 & 
 $ 0 $ 
 & 
 $ 0 $ 
\\
 5 ; 2,2,1 & 
 $ 6 \gamma ^2+12 \gamma +6 $ 
 & 
 $ 8 \gamma ^2+16 \gamma +8 $ 
 & 
 $ 0 $ 
 & 
 $ 0 $ 
\\
 5 ; 2,1,1,1 & 
 $ 0 $ 
 & 
 $ 0 $ 
 & 
 $ 0 $ 
 & 
 $ 0 $ 
\\
 5 ; $1^5$ & 
 $ 0 $ 
 & 
 $ 0 $ 
 & 
 $ 0 $ 
 & 
 $ 0 $ 
\\
 4,1 ; 4,1 & \makecell{
 $ 12 \gamma ^3+24 \gamma ^2 $ \\ $ +24 \gamma +12 $ }
 & 
\makecell{ $ 6 \gamma ^3+24 \gamma ^2 $ \\ $+36 \gamma +18 $ }
 & 
\makecell{ $ 18 \gamma ^3+36 \gamma ^2$ \\ $+18 \gamma $} 
 & 
 \makecell{$ 30 \gamma ^3+90 \gamma ^2 $ \\$+90 \gamma +30 $} 
\\
 4,1 ; 3,2 & 
 $ 12 \gamma ^2+24 \gamma +12 $ 
 & 
 $ 8 \gamma ^2+16 \gamma +8 $ 
 & 
 $ 0 $ 
 & 
 $ 0 $ 
\\
 4,1 ; 3,1,1 & 
 $ 12 \gamma ^2+12 \gamma $ 
 & 
 $ 8 \gamma ^2+8 \gamma $ 
 & 
 $ 12 \gamma ^2+24 \gamma +12 $ 
 & 
 $ 0 $ 
\\
 4,1 ; 2,2,1 & 
 $ 6 \gamma ^2+6 \gamma $ 
 & 
 $ 2 \gamma ^2+2 \gamma $ 
 & 
 $ 6 \gamma ^2+12 \gamma +6 $ 
 & 
 $ 0 $ 
\\
 4,1 ; 2,1,1,1 & 
 $ 6 \gamma +6 $ 
 & 
 $ 4 \gamma +4 $ 
 & 
 $ 0 $ 
 & 
 $ 0 $ 
\\
 4,1 ; $1^5$ & 
 $ 0 $ 
 & 
 $ 0 $ 
 & 
 $ 0 $ 
 & 
 $ 0 $ 
\\
 3,2 ; 3,2 & 
\makecell{ $ 2 \gamma ^3+9 \gamma ^2 $ \\$ +14 \gamma +7 $} 
 & 
\makecell{ $ 4 \gamma ^3+12 \gamma ^2 $ \\ $+16 \gamma +8 $ }
 & 
 $ 8 \gamma ^3+16 \gamma ^2+8 \gamma $ 
 & \makecell{
 $ 20 \gamma ^3+60 \gamma ^2$ \\ $+60 \gamma +20 $ }
\\
 3,2 ; 3,1,1 & 
 $ 0 $ 
 & 
 $ 4 \gamma ^2+4 \gamma $ 
 & 
 $ 2 \gamma ^2+4 \gamma +2 $ 
 & 
 $ 0 $ 
\\
 3,2 ; 2,2,1 & 
 $ 3 \gamma ^2+3 \gamma $ 
 & 
 $ 4 \gamma ^2+4 \gamma $ 
 & 
 $ 6 \gamma ^2+12 \gamma +6 $ 
 & 
 $ 0 $ 
\\
 3,2 ; 2,1,1,1 & 
 $ \gamma +1 $ 
 & 
 $ 4 \gamma +4 $ 
 & 
 $ 0 $ 
 & 
 $ 0 $ 
\\
 3,2 ; $1^5$ & 
 $ 0 $ 
 & 
 $ 0 $ 
 & 
 $ 0 $ 
 & 
 $ 0 $ 
\\
 3,1,1 ; 3,1,1 & 
 $ 2 \gamma ^2+7 \gamma +7 $ 
 & 
 $ 8 \gamma +8 $ 
 & 
 $ 6 \gamma ^2+6 \gamma $ 
 &  \makecell{
 $ 20 \gamma ^2+40 \gamma$ \\ $ +20 $ }
\\
 3,1,1 ; 2,2,1 & 
 $ 6 \gamma +6 $ 
 & 
 $ 4 \gamma +4 $ 
 & 
 $ 0 $ 
 & 
 $ 0 $ 
\\
 3,1,1 ; 2,1,1,1 & 
 $ 3 \gamma $ 
 & 
 $ 0 $ 
 & 
 $ 6 \gamma +6 $ 
 & 
 $ 0 $ 
\\
 3,1,1 ; $1^5$ & 
 $ 1 $ 
 & 
 $ 0 $ 
 & 
 $ 0 $ 
 & 
 $ 0 $ 
\\
 2,2,1 ; 2,2,1 & 
 $ 3 \gamma +3 $ 
 & 
 $ \gamma ^2+2 \gamma +2 $ 
 & 
 $ 3 \gamma ^2+3 \gamma $ 
 & 
\makecell{ $ 15 \gamma ^2+30 \gamma$ \\$ +15 $ }
\\
 2,2,1 ; 2,1,1,1 & 
 $ 0 $ 
 & 
 $ 2 \gamma $ 
 & 
 $ 3 \gamma +3 $ 
 & 
 $ 0 $ 
\\
 2,2,1 ; $1^5$ & 
 $ 0 $ 
 & 
 $ 1 $ 
 & 
 $ 0 $ 
 & 
 $ 0 $ 
\\
 $2,1^3$; $2,1^3$ & 
 $ 3 $ 
 & 
 $ 2 $ 
 & 
 $ \gamma $ 
 & 
 $ 10 \gamma +10 $ 
\\
 $2,1^3$ ; $1^5$ & 
 $ 0 $ 
 & 
 $ 0 $ 
 & 
 $ 1 $ 
 & 
 $ 0 $ 
\\
 $1^5$ ; $1^5$ & 
 $ 0 $ 
 & 
 $ 0 $ 
 & 
 $ 0 $ 
 & 
 $ 1 $ 
\\
 \hline
\end{tabular}
\end{center}
\caption{The Jack connection coefficients and equivariant structure constants for partitions of $5$, part 2.}
\label{JackStructureConstantsAtn=5Part2}
\end{table}

\subsection{A Rudimentary Example of Large Order Stability}
\label{Largen}
To illustrate the idea of the equivariant structure constants stabilizing at large $n$, we provide an example. We consider the structure constants ${c_{2;2}}^\lambda$ for $n=2,3,4,5$ and $n=6$:\footnote{The reader can compare to the discussion of the undeformed case in \cite{Li:2020zwo}. The  rigorous construction in terms of an appropriate infinite order limit of a partial permutation inverse monoid is found in \cite{IvanovKerov} in the non-equivariant case.} 
\begin{eqnarray}
[2] \circ [2] &=& (1+\gamma) [1^2] + \gamma [2]
\nonumber \\
{ [}2,1] \circ [12]&=& 3 (1+\gamma) [1^3] +
\gamma [2,1] + 3 [3]
\nonumber \\
{[}2,1^2] \circ [2,1^2] &=& 6 (1+\gamma) [1^4]
+\gamma [2,1^2]+ 2 [2^2]+ 3 [3,1]
\nonumber \\
{[}2,1^3] \circ [2,1^3] &=& 10 (1+\gamma)[1^5] +\gamma[2,1^3]+ 2 [2^2,1] + 3 [3,1^2]
\nonumber \\
{[}2,1^4] \circ [2,1^4] &=& 15 (1+\gamma) [1^6]
+\gamma [2,1^4] + 2 [2^2,1^2 ] + 3 [3,1^3] \, . \label{alphaNormalizationExample}
\end{eqnarray}
We can introduce  classes renormalized by the number of choices of  $1$ entries that are present at the moment when the class appears on the right hand side among the total number of $1$ entries in the class. Then the classes $[1^k]$ on the right hand side are renormalized by $3,6,10,15$ respectively for $k=3,4,5,6$. In terms of the renormalized classes, the structure constants on the right hand side then stabilize at $n=4$. 
 It would be good to have a generalization of the theorem in the undeformed case \cite{IvanovKerov} that this is always the case. We expect the concept of normalized Jack characters \cite{BurchardtPhD,Sniady} to be useful in this regard.

\bibliographystyle{JHEP}

\begin{thebibliography}{99}

\bibitem{tHooft:1993dmi}
G.~'t Hooft,
``Dimensional reduction in quantum gravity,''
Conf. Proc. C \textbf{930308} (1993), 284-296
[arXiv:gr-qc/9310026 [gr-qc]].

\bibitem{Susskind:1994vu}
L.~Susskind,
``The World as a hologram,''
J. Math. Phys. \textbf{36} (1995), 6377-6396
doi:10.1063/1.531249
[arXiv:hep-th/9409089 [hep-th]].

\bibitem{Maldacena:1997re}
J.~M.~Maldacena,
``The Large $N$ limit of superconformal field theories and supergravity,''
Adv. Theor. Math. Phys. \textbf{2} (1998), 231-252
doi:10.4310/ATMP.1998.v2.n2.a1
[arXiv:hep-th/9711200 [hep-th]].



\bibitem{Eberhardt:2019ywk}
L.~Eberhardt, M.~R.~Gaberdiel and R.~Gopakumar,
``Deriving the AdS$_{3}$/CFT$_{2}$ correspondence,''
JHEP \textbf{02} (2020), 136
doi:10.1007/JHEP02(2020)136
[arXiv:1911.00378 [hep-th]].




\bibitem{Li:2020zwo}
S.~Li and J.~Troost,
``The Topological Symmetric Orbifold,''
JHEP \textbf{10} (2020), 201
doi:10.1007/JHEP10(2020)201
[arXiv:2006.09346 [hep-th]].





\bibitem{BryanGraber}
J.~Bryan, T.~Graber,  ``The crepant resolution conjecture,'' In Proc. Sympos. Pure Math (Vol. 80, pp. 23-42) (2009).

\bibitem{BryanPandharipande}

J.~Bryan, R.~Pandharipande,  ``The local Gromov-Witten theory of curves,'' Journal of the American Mathematical Society, 21(1) (2008), 101-136. 


\bibitem{OPQuantumCohomology}
A.~Okounkov, R.~Pandharipande, ``Quantum cohomology of the Hilbert scheme of points in the plane,'' Inventiones mathematicae, 179(3), 523-557 (2010).



\bibitem{OP1}
A.~Okounkov and R.~Pandharipande,
``Gromov-Witten theory, Hurwitz theory, and completed cycles,''
Ann. Math. \textbf{163} (2006), 517-560
doi:10.4007/annals.2006.163.517
[arXiv:math/0204305 [math]].
\bibitem{OP2}
A.~Okounkov and R.~Pandharipande,
``The Equivariant Gromov-Witten theory of P1,''
[arXiv:math/0207233 [math.AG]].
%
\bibitem{OP3}
A.~Okounkov and R.~Pandharipande,
``Virasoro constraints for target \- curves,''
[arXiv: \- math/0308097 \- [math.AG]].



\bibitem{LS} M.~Lehn, C.~Sorger,  ``Symmetric groups and the cup product on the cohomology of Hilbert schemes,'' Duke Mathematical Journal, 110(2), 345-357  (2001).


\bibitem{Ashok:2023mow}
S.~K.~Ashok and J.~Troost,
``The chiral ring of a symmetric orbifold and its large N limit,''
JHEP \textbf{08} (2023), 004
doi:10.1007/JHEP08(2023)004
[arXiv:2303.09308 [hep-th]].

\bibitem{Pakman:2009ab}
A.~Pakman, L.~Rastelli and S.~S.~Razamat,
``Extremal Correlators and Hurwitz Numbers in Symmetric Product Orbifolds,''
Phys. Rev. D \textbf{80} (2009), 086009
doi:\-10.1103/\-PhysRevD.80.086009
[arXiv:0905.3451 [hep-th]].

\bibitem{Gaberdiel:2007vu}
M.~R.~Gaberdiel and I.~Kirsch,
``Worldsheet correlators in AdS(3)/CFT(2),''
JHEP \textbf{04} (2007), 050
doi:10.1088/1126-6708/2007/04/050
[arXiv:hep-th/0703001 [hep-th]].

\bibitem{Dabholkar:2007ey}
A.~Dabholkar and A.~Pakman,
``Exact chiral ring of AdS(3) / CFT(2),''
Adv. Theor. Math. Phys. \textbf{13} (2009) no.2, 409-462
doi:10.4310/ATMP.2009.v13.n2.a2
[arXiv:hep-th/0703022 [hep-th]].

\bibitem{Nekrasov:2002qd}
N.~A.~Nekrasov,
``Seiberg-Witten prepotential from instanton counting,''
Adv. Theor. Math. Phys. \textbf{7} (2003) no.5, 831-864
doi:10.4310/ATMP.2003.v7.n5.a4
[arXiv:\-hep-th/\-0206161 [hep-th]].

\bibitem{Nekrasov:2010ka}
N.~Nekrasov and E.~Witten,
``The Omega Deformation, Branes, Integrability, and Liouville Theory,''
JHEP \textbf{09} (2010), 092
doi:10.1007/JHEP09(2010)092
[arXiv:1002.0888 [hep-th]].

\bibitem{Aganagic:2004js}
M.~Aganagic, H.~Ooguri, N.~Saulina and C.~Vafa,
``Black holes, q-deformed 2d Yang-Mills, and non-perturbative topological strings,''
Nucl. Phys. B \textbf{715} (2005), 304-348
doi:10.1016/j.nuclphysb.2005.02.035
[arXiv:hep-th/0411280 [hep-th]].

\bibitem{Benizri:2024mpx}
L.~Benizri and J.~Troost,
``Symmetric group gauge theories and simple gauge/string du\-ali\-ties,''
J. Phys. A \textbf{57} (2024) no.50, 505401
doi:10.1088/\-1751-8121/\-ad92ce
[arXiv:\-2404.12543 [hep-th]].



\bibitem{Eberhardt:2021vsx}
L.~Eberhardt,
``A perturbative CFT dual for pure NS{\textendash}NS AdS$_{3}$ strings,''
J. Phys. A \textbf{55} (2022) no.6, 064001
doi:10.1088/1751-8121/ac47b2
[arXiv:2110.07535 [hep-th]].




\bibitem{Troost:2026pmx}
J.~Troost,
``The Open/Closed Gromov-Witten/Hurwitz Correspondence and Localized World Sheets for Completed Cycles,''
J. Phys. A: Math. Theor. https://doi.org/10.1088/1751-8121/ae7f53, 
[arXiv:2602.12786 [hep-th]] .



\bibitem{Witten:1989ig}
E.~Witten,
``On the Structure of the Topological Phase of Two-dimensional Gravity,''
Nucl. Phys. B \textbf{340} (1990), 281-332
doi:10.1016/0550-3213(90)90449-N

\bibitem{Faber:1998gsw}
C.Faber and R.Pandharipande,
``Hodge integrals and Gromov-Witten theory,''
In\-ven\-tiones mathematicae \textbf{139} (2000) no.1, 173-199
doi:\-10.1007\-/s002229900028
[arXiv:\-math/\-9810173 [math.AG]].

\bibitem{Dijkgraaf:1996xw}
R.~Dijkgraaf, G.~Moore, E.~Verlinde and H.~Verlinde,
``Elliptic genera of symmetric products and second quantized strings,''
Commun. Math. Phys. \textbf{185} (1997), 197-209
doi:10.1007/s002200050087
[arXiv:hep-th/9608096 [hep-th]].



\bibitem{Okounkov}
A.~Okounkov, ``The uses of random partitions,'' In XIVth International Congress on Mathematical Physics (pp. 379-403) (2006). 

\bibitem{Morozov:2013rma}
A.~Morozov and A.~Smirnov,
``Towards the Proof of AGT Relations with the Help of the Generalized Jack Polynomials,''
Lett. Math. Phys. \textbf{104} (2014) no.5, 585-612
doi:10.1007/s11005-014-0681-6
[arXiv:1307.2576 [hep-th]].

\bibitem{Fogarty}
J.~Fogarty,  ``Algebraic families on an algebraic surface,'' American Journal of Mathematics, 90(2), 511-521 (1968).



\bibitem{Witten:1982df}
E.~Witten,
``Constraints on Supersymmetry Breaking,''
Nucl. Phys. B \textbf{202} (1982), 253
doi:10.1016/0550-3213(82)90071-2.


\bibitem{Witten:1982im}
E.~Witten,
``Supersymmetry and Morse theory,''
J. Diff. Geom. \textbf{17} (1982) no.4, 661-692.




\bibitem{Morse}
M.~Morse, ``The Calculus of Variations in the Large,'' American Mathematical Society Colloquium Publication. Vol. 18. New York (1934).

\bibitem{NakajimaLectures}
H.~Nakajima, ``Lectures on Hilbert schemes of points on surfaces,'' (No. 18). American Mathematical Soc. (1999).



\bibitem{ES}
G.~Ellingsrud, S.~Strømme, ``On the homology of the Hilbert scheme of points in the
plane,'' Invent. Math. 87 (1987), 343-3.


\bibitem{NakajimaJack}
H.~Nakajima, ``Jack polynomials and Hilbert schemes of points on surfaces,'' arXiv preprint alg-geom/9610021.

\bibitem{Vasserot}
E.~Vasserot, ``Sur l'anneau de cohomologie du schéma de Hilbert de C2,'' Comptes Rendus de l'Académie des Sciences-Series I-Mathematics, 332(1), 7-12  (2001). 

\bibitem{LQW}
W.~Li,  Z.~Qin,  W.~Wang, ``The cohomology rings of Hilbert schemes via Jack polynomials,'' arXiv preprint math/0411255.


\bibitem{NakajimaMoreLectures}
H.~Nakajima, ``More lectures on Hilbert schemes of points on surfaces,'' arXiv preprint arXiv:1401.6782.



\bibitem{AB}
M.~Atiyah and R.~Bott. ``The Moment Map and Equivariant Cohomology,'' Topology {\bf 23.1} (1984), 1-28.

\bibitem{Macdonald}
I.~Macdonald, ``Symmetric functions and Hall polynomials,'' Oxford university press (1998). 


\bibitem{Grojnowski}
I.~Grojnowski, ``Instantons and affine algebras I: the Hilbert scheme and vertex operators,'' arXiv preprint alg-geom/9506020.  

%


\bibitem{HSS}
P.~Hanlon, R.~Stanley, J.~Stembridge, ``Some combinatorial aspects of the spectra of normally distributed random matrices,'' Contemp. Math, 138, 151-174 (1992). 

\bibitem{Burnside}
W.~Burnside, ``Theory of Groups of Finite Order,'' 2nd Edition, Dover Publications, 1955.


\bibitem{FJW}
I.~Frenkel, N.~Jing, W.~Wang,  ``Vertex representations via finite groups and the McKay correspondence,'' International Mathematics Research Notices, 2000(4), 195-222 (2000).


\bibitem{FarahatHigman}
H.~Farahat, G.~Higman, ``The centres of symmetric group rings,'' Proceedings of the Royal Society of London. Series A. Mathematical and Physical Sciences, 250(1261), 212-221  (1959). 



\bibitem{GouldenJackson}
I.~Goulden and D.~Jackson, ``Connection coefficients, matchings, maps and combinatorial conjectures for Jack symmetric functions,'' Transactions of the American Mathematical Society 348.3 (1996): 873-892.


\bibitem{IvanovKerov}
V.~Ivanov, S.~Kerov, ``The algebra of conjugacy classes in symmetric groups and partial permutations,'' Journal of Mathematical Sciences 107.5, 4212-4230 (2001).


\bibitem{Wang}
W.~Wang, ``The Farahat–Higman ring of wreath products and Hilbert schemes,'' Advances in Mathematics, 187(2), 417-446  (2004). 

\bibitem{DolegaFeray}

M.~Doł{e}ga, V.~Féray, ``Gaussian fluctuations of Young diagrams and structure constants of Jack characters,'' Duke Mathematical Journal, 165(7), 1193-1282 (2016).


\bibitem{Vassilieva}
E.~Vassilieva, ``Polynomial properties of Jack connection coefficients and generalization of a result by Dénes,''  Journal of Algebraic Combinatorics, 42(1), 51-71 (2015).


\bibitem{BenDali}
H.~Ben Dali, ``Integrality in the Matching-Jack conjecture and the Farahat-Higman algebra,''  arXiv:2203.14879.


\bibitem{Burchardt}
A.~Burchardt, ``The top-degree part in the Matchings-Jack Conjecture,'' arXiv preprint arXiv:1803.09330.




\bibitem{BurchardtPhD}
A.~Burchardt, ``Structure constants of Jack characters,'' PhD dissertation, Adam Mickiewicz University, Poznan 2018.



\bibitem{Sniady}
P.~Sniady, ``Structure coefficients for Jack characters: approximate factorization property,'' arXiv preprint arXiv:1603.04268.


\bibitem{FW}
I.~Frenkel, W.~Wang, ``Virasoro algebra and wreath product convolution,'' Journal of Algebra, 242(2), 656-671  (2001).

\bibitem{Dijkgraaf:1996iy}
R.~Dijkgraaf,
``Chiral deformations of conformal field theories,''
Nucl. Phys. B \textbf{493} (1997), 588-612
doi:10.1016/S0550-3213(97)00153-3
[arXiv:hep-th/9609022 [hep-th]].

\bibitem{Douglas:1993wy}
M.~Douglas,
``Conformal field theory techniques in large N Yang-Mills theory,''
[arXiv:hep-th/9311130 [hep-th]].


\bibitem{GouldenJacksonMacdonald}
I.~Goulden, D.~Jackson, ``Symmetrical Functions and Macdonald's Result for Top Connexion Coefficients in the Symmetrical Group,'' Journal of Algebra, 166(2), 364-378  (1994). 

\bibitem{Alexandersson}
The Mathematica Package Symmetric Functions by Per Alexandersson, downloaded from the URL
https://github.com/PerAlexandersson/Mathematica-packages on 22/11/2025.








\end{thebibliography}

\end{document}